\documentclass[preprint2]{aastex}

\newcommand{\be}{\begin{equation}}
\newcommand{\ee}{\end{equation}}

\newcommand{\ax}{$\alpha_{\rm X}$}

\newcommand{\aox}{$\alpha_{\rm ox}$}

\newcommand{\msun}{$M_{\odot}$}

\newcommand{\plm}{$\pm$}

\newcommand{\swift}{{\it Swift}}
\newcommand{\suzaku}{{\it Suzaku}}
\newcommand{\xmm}{{\it XMM-Newton}}

\def\et{{et al.\ }}

\def\mrk{{Mkn~335}}

\shorttitle{XMM observation of Mrk 335}
\shortauthors{Grupe et al.}

\begin{document}


\def\etal{{\it et\thinspace al.}\ }
\def\alp{{$\alpha$}\ }
\def\al2{{$\alpha^2$}\ }

%
%
%


\title{\xmm\ observations of the Narrow-Line Seyfert 1 galaxy Mrk 335 
in an historical low X-ray flux state 
}


\author{Dirk Grupe\altaffilmark{1}, 
\email{grupe@astro.psu.edu}
Stefanie Komossa\altaffilmark{2},
Luigi C. Gallo\altaffilmark{3},
Andrew C. Fabian\altaffilmark{4},
Josefin Larsson\altaffilmark{4},
Anil K. Pradhan\altaffilmark{5},
Dawei Xu\altaffilmark{6},
Giovanni Miniutti\altaffilmark{4}
}

\altaffiltext{1}{Department of Astronomy and Astrophysics, Pennsylvania State
University, 525 Davey Lab, University Park, PA 16802} 

\altaffiltext{2}{Max-Planck-Institut f\"ur extraterrestrische Physik, Giessenbachstr., D-85748 Garching,
Germany; email: skomossa@mpe.mpg.de}

\altaffiltext{3}{Department of Astronomy and Physics, Saint Mary's University, Halifax, NS
B3H 3C3, Canada;
email: lgallo@ap.stmarys.ca}

\altaffiltext{4}{Institute of Astronomy, Madingley Road, Cambridge, CB3 0HA, UK}

\altaffiltext{5}{Department of Astronomy, The Ohio State University, 140 W 18th
Av, Columbus, OH; pradhan@astronomy.ohio-state.edu}

\altaffiltext{6}{National Astronomical Observatories, Chinese Academy of Science, Beijing 100012, China; 
dwxu@bao.ac.cn}




\begin{abstract}
We report the discovery of strong soft X-ray emission lines 
and a hard continuum above 2 keV
in the Narrow-Line Seyfert 1 galaxy Mrk 335
during an extremely low X-ray flux state.
Mrk 335 was observed for 22 ks by \xmm\
in July 2007 as a Target of Opportunity to examine it in its
X-ray low-flux state, which was discovered with \swift. 
Long-term light curves suggest that this is the lowest flux state this AGN has ever
been seen in. 
However, Mrk 335 is still sufficiently bright that 
its X-ray properties can be studied in
detail.
The X-ray continuum spectrum is very complex and requires several components to
model.  Statistically, partial covering and blurred reflection models work well.
We confirm the presence of a strong  narrow Fe line at 6.4 keV. 
High-resolution spectroscopy with the \xmm-RGS reveals strong, soft X-ray emission
lines not detected in previous, higher signal-to-noise, \xmm\
observations, such as: highly
ionized Fe lines, O VII, Ne IX and Mg XI lines.  
The optical/UV fluxes are 
 similar to those previously measured with \swift.
Optical spectroscopy taken in 2007 September do not show any changes to optical spectra obtained 
8 years earlier. 
\end{abstract}


\section{Introduction}

The Narrow-Line Seyfert 1 galaxy \citep[NLS1; ][]{oster85} Mrk 335
($\alpha_{2000}$ = $00^{\rm h} 06^{\rm m} 19.^{\rm s}5$, 
$\delta_{2000}$ = $+20^{\circ} 12' 11 \farcs 0$, z=0.026) was discovered as a
bright X-ray AGN  with UHURU \citep{tananbaum78} in 1971. Consequently it has been
a target of almost all X-ray observatories including: 
EINSTEIN, EXOSAT, GINGA, ROSAT,
ASCA, XMM and Suzaku 
\citep{halpern82,pounds87, nandra94, grupe01, george00,gondoin02,
longinotti07a,longinotti07b,oneill07, larsson07}. The X-ray spectrum of Mrk 335 appears to
be rather complicated and cannot be modeled by a single power law spectrum.
Possible interpretations for the X-ray spectrum include: 
absorption by ionized gas
\citep[e.g. ][]{nandra94, lei99b}, partial-covering \citep[e.g.][]{tanaka05}, and 
X-ray reflection on the accretion
disk \citep[e.g. ][]{ballantyne01, crummy06, longinotti07a, oneill07, larsson07}. Support for the
reflection interpretation comes from the presence of a broadened and asymmetric Fe
K$\alpha$ line \citep{longinotti07a, oneill07}. Such lines are predicted as
a consequence of reflection of the primary power law on the
accretion disk close to the black hole \citep{fabian89}
and are seen
in a growing number of AGN \citep[see ][and references therein]{nandra07}.
Mrk 335 is a highly variable X-ray source that typically varies by factors of
2 to 4 on hourly time scales \citep[e.g. ][]{turner93,oneill07}. 

However, when
Mrk 335 was observed with the Gamma-Ray Burst Explorer Mission
\swift\ \citep{gehrels04}  in 2007 May and June \citep{grupe07c}
it appeared significantly 
fainter in its X-ray flux 
by factors of more than 30 compared to the 2006 \xmm\, and Suzaku observations 
\citep[][]{longinotti07b, oneill07, larsson07}. This drop in X-rays coincided with substantial
changes in the X-ray spectrum.  Mrk 335 showed a very hard X-ray component at energies above 2 keV,
very similar to what has been seen in other NLS1 galaxies (e.g. Gallo 2006) and specifically in NGC 4051
\citep{guainazzi98, ponti06}. 

As discussed in \citet{grupe07c}, one
possible explanation for the sudden
drop in the X-ray flux of Mrk 335 is a partial covering 
absorber that intersects our line-of-sight with the X-ray emitting zone. 
\citet{grupe07c} found that when correcting for 
this absorber the intrinsic X-ray
flux of Mrk 335 between the high- and low-flux states varied only by factors of
4-6. 

In order to investigate the
nature of the low-state in more detail we initiated a Target-of-Opportunity (ToO)
observation with \xmm, which was executed on 2007 July 10. Prior to this, 
Mrk 335 was observed by \xmm\ in 2000 and 2006 
\citep{gondoin02, crummy06, longinotti07a, oneill07}.  Of interest with Mrk 335 is the broad component
in the spectrum associated with the red-wing of a relativistic iron line. This wing 
is not required if the spectrum is modified by a partial covering absorber. 
The brightness of Mrk 335 provides us with the
rare opportunity to study an AGN in its low-state, but still with sufficient signal
to obtain well-exposed spectra in a relatively short observation. 
In this paper we report the \xmm\ ToO observation. 

The paper is organized as follows: in \S 2 we present the \xmm\ 
and \swift\ observations and
describe the data reduction, in \S 3 we present the results that are
discussed in \S 4.
Throughout the paper spectral indeces are denoted as energy spectral indeces
with
$F_{\nu} \propto \nu^{-\alpha}$. Luminosities are calculated assuming a $\Lambda$CDM
cosmology with $\Omega_{\rm M}$=0.27, $\Omega_{\Lambda}$=0.73 and a Hubble
constant of $H_0$=75 km s$^{-1}$ Mpc$^{-1}$ corresponding to 
 a luminosity distance D=105 Mpc.
All errors are 90\% confidence unless stated otherwise.

\section{\label{observe} Observations and data reduction}

\subsection{\xmm\ and \swift\ Observations}

\xmm\ \citep{jansen01}
observed Mrk 335 on 2007 July 10 for a total of 22 ks (ObsID  0510010701).
A summary of the
observations with each of the instruments on-board \xmm\ is given in  
Table\,\ref{obs_log}. The European Photon Imaging Camera (EPIC) pn
\citep{strueder01} was operated in Large Window  mode with the thin
filter. This combination was chosen to avoid pileup in case the AGN 
re-brightened. The two EPIC MOS \citep{turner01} were both operated in Full-Frame
mode with the medium filters. 
High-resolution X-ray spectroscopy was performed
using the two Reflection Grating Spectrometers \citep[RGS; ][]{denherder01}
on-board \xmm.  Optical photometry was performed in 5 filters with 
the Optical Monitor \citep[OM; ][]{mason01}.  The data are used 
to measure the optical-to-X-ray spectral energy distribution of Mkn
335 during the \xmm\ observation. Due to slew problems at the beginning of the
observations, V-filter observations were not obtained. All OM observations were
performed in a science-user defined configuration with a 7$^{'}\times 7^{'}$
observing window.

The \xmm\  data were processed in the standard way using the 
XMMSAS version {\it xmmsas\_20060628\_1801-7.0.0}. 
The EPIC pn data were checked for episodes of high particle background.
The background count rate was found low throughout the observation.
The source X-ray photons in the EPIC pn 
were selected in a circular region with a radius of 1$^{'}$.  Likewise, 
background photons were selected from a nearby, source-free region 
with the same radius. Only single and double events ({\tt PATTERN.le.4}) 
and single to quadruple events ({\tt PATTERN.le.12}) 
were selected
 for the pn and MOS data, respectively.
The spectra were rebinned with {\it grppha} version 3.0.0 with 50 photons 
per bin. The redistribution matrices and the auxiliary response files were created by the
XMMSAS tasks {\it rmfgen} and {\it arfgen}, respectively. 
RGS spectra and response matrixes were created by the standard RGS XMMSAS tool
{\it rgsproc}. The RGS spectra were rebinned with 10 photons per bin using
{\it grppha}. Spectral fits to the EPIC pn and MOS, and RGS spectra were performed with XSPEC
version 12.3.1x \citep{arnaud96}.
The OM data were processed with the XMMSAS task {\it omichain}. The magnitudes and fluxes
of Mrk 335 were taken from the source lists created by the {\it omichain} task. 
For the count rate to flux conversion we used the conversion factors given in
the OM Calibration document {\tt XMM-SOC-CAL-TN-0019}.

The \swift\ X-ray Telescope \citep[XRT; ][]{burrows05} and UV/Optical Telescope
\citep[UVOT; ][]{roming05} 2007 May and June 
observations and the data reductions are 
described in \citet{grupe07c}. In order to compare the photometry in the OM with the 
\swift\ UVOT we selected 5 field stars with similar brightness in V as Mrk 335. 
The measured (uncorrected) magnitudes of these stars are listed in
Table\,\ref{ref_stars}. We confirm the earlier result by \citet{grupe07c} 
that in  U the magnitudes in the OM and UVOT
agree perfectly. However, the OM magnitude had to be adjusted by $-$0.10
mag in B and $+$0.30 in the UVM2 filter. In addition to the 2000 \xmm\ observation,
during the 2007 \xmm\ ToO observation we also observed in UV W1 and UV W2 as listed
in Table\,\ref{obs_log}. In these filters the OM magnitudes have to be adjusted, 
compared to the \swift\ UVOT, by
$+$0.78 mag and $+$0.15 mag for the UVW1 and UVW2 filters, respectively. 
The 5 stars also show that the OM and UVOT detectors are stable and 
no variability is seen in these 5 stars. This suggests
that any variability detected in the Optical/UV in Mrk 335 is real. 
A summary of the UVOT calibration has recently been published by \citet{poole07}.

\subsection{Optical Spectroscopic Observation}

Optical spectra of Mrk 335 were obtained with the 2.16\,m telescope at the
Xinglong Observatory, National Astronomical Observatories of
China (NAOC) on 2007 September 07. Two spectra of Mrk 335 were
taken with 30 and 60 min exposures during clear weather
conditions and 2$^{''}$ seeing. The spectrograph was
equipped with a 600 lines per mm grating and the
2$^{''}$ slit resulting in a resolution of 5\AA\ in first order,
covering the wavelength range between 4377 to 7120\AA.
The optical data were reduced and analyzed in a standard way
with the ESO Munich Image Data Analysis System (MIDAS, version
06FEBpl1.0). All line measurements were performed on the coadded spectrum after
subtracting the \citet{bor92} I Zw 1 Fe II template as described in
\citet{grupe04}. For the H$\beta$ line we subtracted the narrow component by
constructing a narrow-line template from the [OIII]5007 line as described in
\citet{grupe04}.

\section{\label{results} Results}

\subsection{Temporal Behavior}

\subsubsection{Long-Term Light curves}

\begin{figure}
\epsscale{1.0}
\plotone{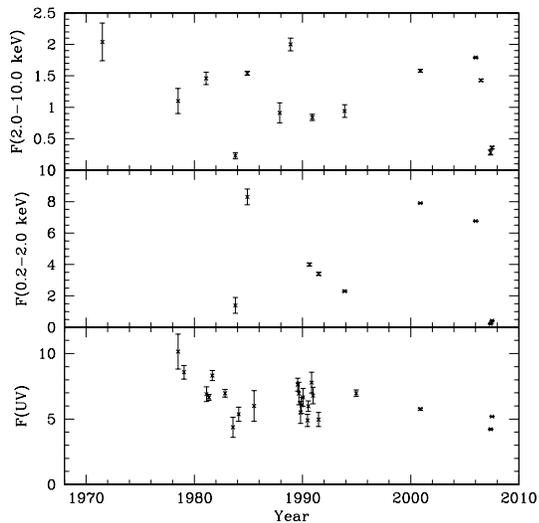}
\caption{\label{mkn335_lc} Long-term light curves of Mrk 335. The upper panel shows the
 2-10 keV light curve starting with the first X-ray detection of
Mkn335 by UHURU in 1971, the middle panel displays the 0.2-2.0 keV flux light curve starting at the EXOSAT 
observation in November 1983.  The
lower panel displays the UV  flux light curve measured with
 IUE and  HST at 2256\AA, and \xmm-OM and \swift-UVOT using the M2 filters. 
 Note that these UV
data are not corrected for Galactic reddening. The \xmm\ data are adjusted to the \swift\ UVOT data.
All fluxes are given in units of $10^{-14}$ W m$^{-2}$ (=$10^{-11}$ ergs s$^{-1}$ cm$^{-2}$) 
as listed in Table\,\ref{lc_summary}.
}
\end{figure}

Figure\,\ref{mkn335_lc} displays the observed 2.0-10.0 keV, 0.2-2.0 keV, and UV (2256\AA) 
long-term
light curves of Mrk 335 covering the time between 1971
and 2007 July.  
The X-ray fluxes and references
are listed in Table\,\ref{lc_summary}.
The UV fluxes were derived from the UV IUE and HST database by \citet{dunn06}.
To remain consistent with \citet{dunn06}, the \xmm\ OM and \swift\ UVOT light curves were created in the UV M2 filters, 
since its central wavelength is closest to 2256\AA.
As noticed by
\citet{pounds87}, the EXOSAT data of
Mrk 335 suggest that it was in an unusual X-ray low-state in 1983. This seems coincident
with a minimum in the UV light curve as well.
However, about a year later in December 1984, Mrk 335 was back in a typical high state.
The plots in Figure\,\ref{mkn335_lc} demonstrate that from 2007 May to July 
Mrk 335 was seen in an historic X-ray low-state.
This drop happened in less than
one year since Mrk 335 was last observed in June 2006 with \suzaku\ \citep{larsson07}.
In order to determine 
when and how fast this recent flux drop occurred, the RXTE ASM 
light curve  was examined. 
The month-to-month 
ASM light curve indicates that Mrk 335 has been
variable in X-rays over the last decade. However, Mrk 335 
 is too faint to constrain the occurrence of the most recent flux drop 
 with the ASM.
Mrk 335 also appears to be fainter in the UV in 2007 compared with previous
epochs. Table\,\ref{mkn335_uv} lists the \xmm\ OM and \swift\ UVOT measurements.
During the 2007 May 17 \swift\ observation Mrk 335 appears to be fainter 
in the UV by
about 0.2 mag compared to the \swift\ and \xmm\ observations before and after.

\subsubsection{\label{short-term} Short-term Variability}

\begin{figure}
\epsscale{0.75}
\plottwo{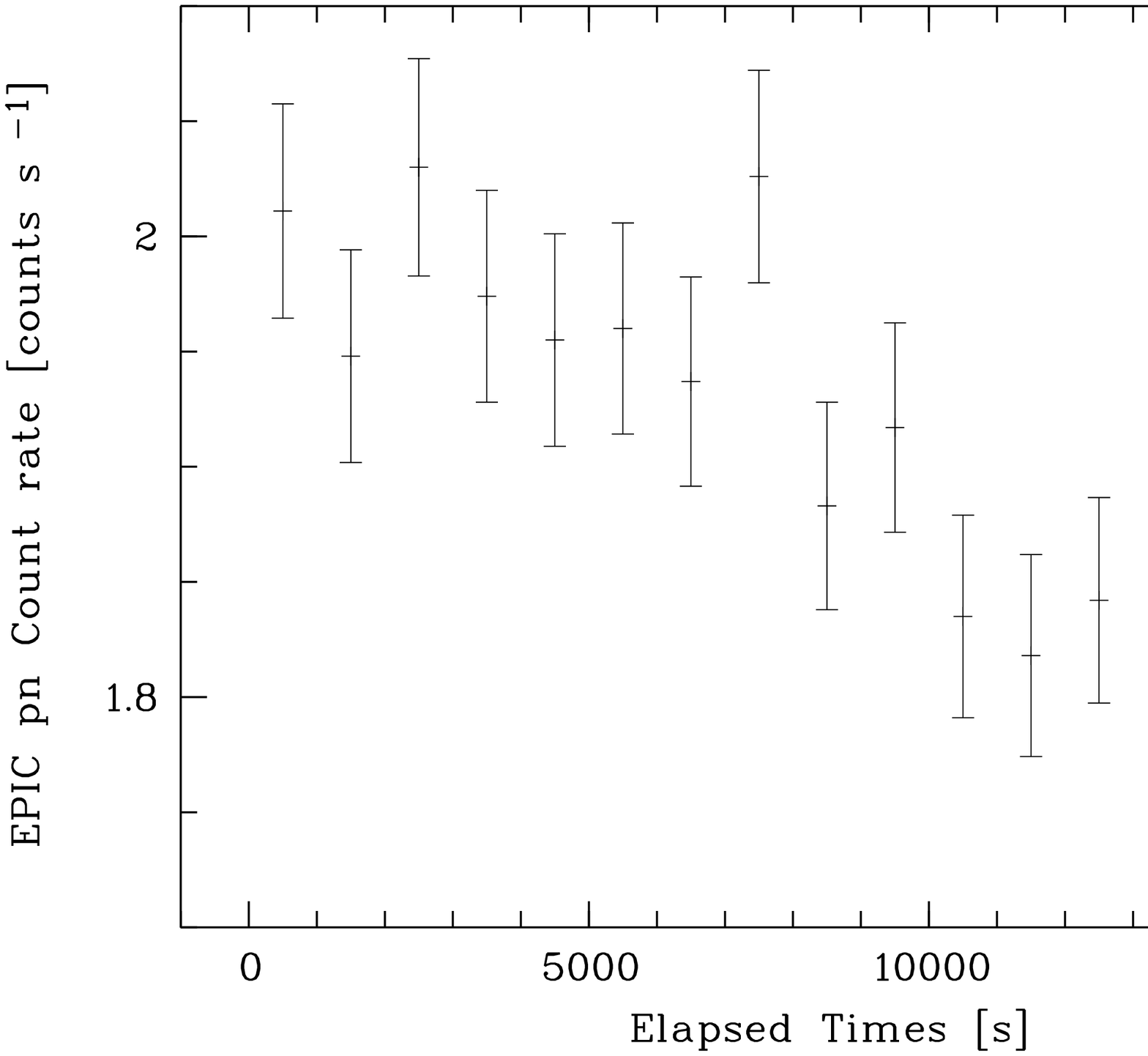}{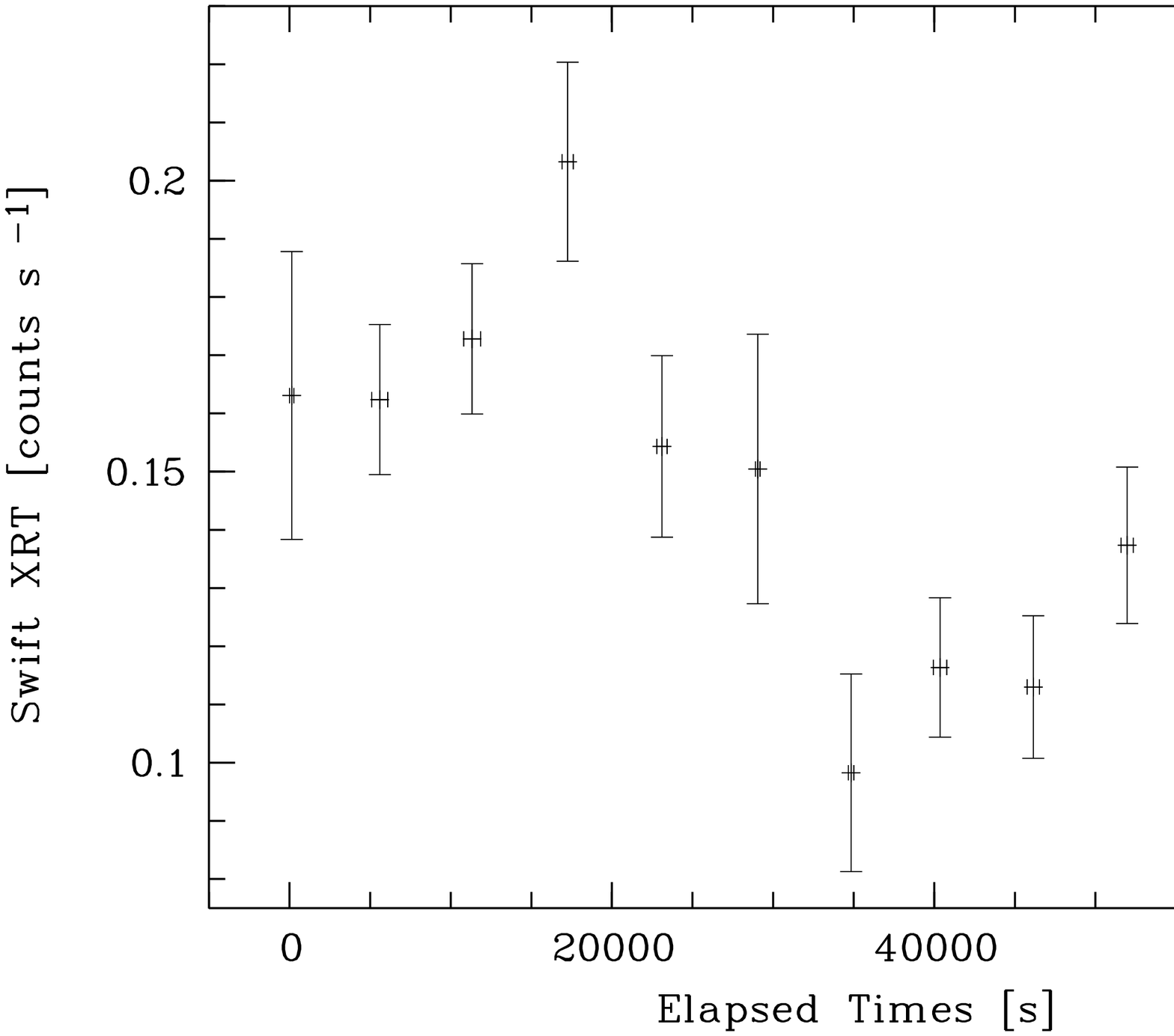}
\caption{\label{mkn335_pn_lc} 2007 \xmm\ EPIC pn light curve 
in the 0.2-10.0 keV band (left) and \swift\ XRT 
0.3-10.0 keV light curve of the
second \swift\ observation on 2007 May 25 \citep{grupe07c} on the right panel.
The starting times of the observations are 2007 July 10 17:09 UT and 2007 May 25 00:01 UT, respectively.
}
\end{figure}

The left panel of Figure\,\ref{mkn335_pn_lc} shows the 2007 July
\xmm\ EPIC pn 
light curve in the 0.2-10.0 keV band. 
During the 16.8 ks exposure
the AGN showed a steady decay in flux by about 15\%. 
The light curve is relatively quiescent compared to the high-flux state
variability seen in the 2006 \xmm\ and {\it Suzaku} light curves 
\citep[][respectively]{oneill07, larsson07}.
The variability is in agreement with the 
variability - flux dependence reported by \citet{uttley01}.
However, the 2007
July \xmm\ observation was rather short (17 ks). 
During the low-state \swift\ observation on 2007 May 25, which covers a time period of
70 ks,
Mrk 335 is quite variable by a factor of about 2 within 20 ks
(right panel in Figure\,\ref{mkn335_pn_lc}). In both cases the AGN is variable
and deviates significantly from a constant value. $\chi^2$ tests show that the
$\chi^2/\nu$ is 49/15 and 55/11 for the \xmm\ pn and \swift\ XRT light curves,
respectively, assuming the mean count rates as constant values.

\begin{figure}
\epsscale{1.0}
\plotone{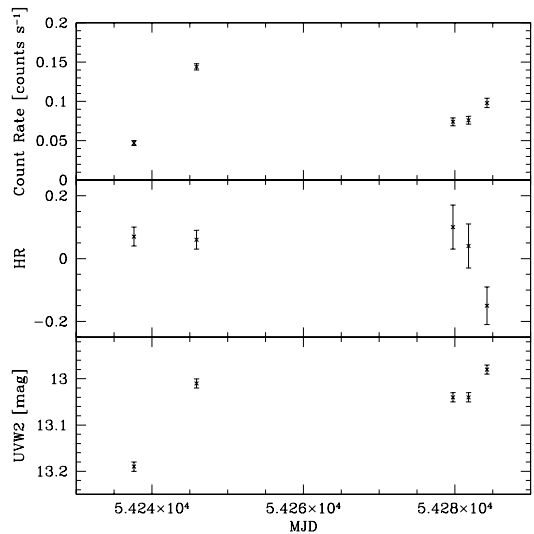}
\caption{\label{mkn335_xrt_uvot_lc} \swift\ XRT and UVOT-UVW2 light curves of Mkn335 between
2007 May 17--June 02.  The upper panel shows the XRT count
rate, the middle panel the XRT hardness ratio (see \S\,\ref{short-term}),
and the lower panel displays the UVOT-UVW2 light curve. 
}\end{figure}

In Figure\,\ref{mkn335_xrt_uvot_lc} 
the XRT and UVW2 light curves for all  \swift\ segments between 2007 May 17 
and July 02 are displayed
(the magnitudes of 
segments 001-005 are listed in Table\,\ref{mkn335_uv}). 
The light curves show that Mrk 335 is quite variable on intermediate
time scales of days to weeks. Note that the UVW2 light curve follows 
the XRT count rate suggesting some connection between the Optical/UV 
and X-ray variability.
The hardness
ratio\footnote{The hardness ratio is defined as $HR=(H-S)/(H+S)$ (where S and H are
the number of photons in the 0.3-1.0 keV and 1.0-10.0 keV bands, respectively).} 
variability curve may suggest some spectral variability in the X-ray band
(middle panel of Figure\,\ref{mkn335_xrt_uvot_lc}), however a $\chi^2$ test
shows that this is not significant considering the uncertainties.

\subsection{X-ray Spectral analysis}

\begin{figure}
\epsscale{0.7}
\plotone{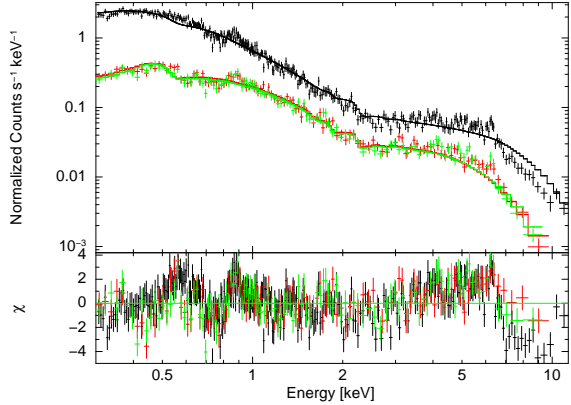}
\caption{\label{mkn335_xray_epic} \xmm\ EPIC pn 0.3-12.0 keV (black)
and MOS 0.3-10.0 keV (red + green)
spectra of Mrk 335 fitted with a
broken power law with $\alpha_{\rm X,soft}$=1.89, $\alpha_{\rm X,hard}=-0.19$, 
$E_{\rm break}$=1.73 keV as listed in Table\,\ref{xray_res}. 
The absorption column density is fixed to the Galactic value
\citep[$3.96 \times 10^{20}$ cm$^{-2}$; ][]{dic90}.
}
\end{figure}

\begin{figure}
\epsscale{0.8}
\plottwo{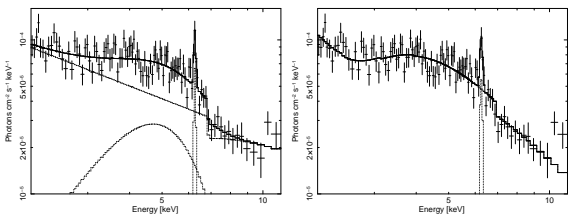}{f5b.ps}
\caption{\label{mkn335_broad_fe} \xmm\ EPIC pn 2.0-12.0 keV
spectrum of Mrk 335 with a power law and a broad Fe line at 4.9 keV (left panel)
and a power law with partial covering absorber (right panel).
}
\end{figure}

Table\,\ref{xray_res} summarizes the results from the X-ray spectral
analysis in the 0.3-12.0 keV range of
the 2007 \xmm\ ToO (low-flux state) and 2006 \xmm\ GO (high-flux state) observations. 
The 2007
EPIC pn and MOS spectra cannot be fitted by a single absorbed
power law model. This is similar to the previous \xmm\ and \swift\ observations \citep{grupe07c} which cannot be fitted by a single power law
model either.
Applying a broken power law model to the pn and MOS data
 significantly improves the fit,
but as shown in Figure\,\ref{mkn335_xray_epic} the X-ray spectrum during the 2007 \xmm\ ToO
observation is quite complicated and requires further components.  
There are strong residuals at energies below 1 keV and between 3-7 keV.
The residuals at lower energies can be associated with strong
X-ray emission lines found in the RGS spectra of Mrk 335 (see \S\,\ref{rgs}). 
Figure\,\ref{mkn335_xray_epic} also shows that the pn and MOS data are consistent with
each other. 
Due to some internal processing problems, MOS data were not made available to us in time
to be handled in the full analysis. 
In order to represent soft X-ray emission lines we added two
Gaussian profiles at 0.5 and 0.9 keV. These Gaussians do not represent specific lines, but are used to mimic line
emission and improve the fit. The emission feature at 6.4 keV is associated with near-neutral Fe K$\alpha$ emission (e.g Fe I-XVII),
and is modeled with a single, narrow Gaussian.
The addition of these three Gaussian profiles to the broken power law model improves the fit,
but significant deviations remain in the 2-6 keV range. These residuals can be improved with 
an extremely broad Gaussian with a peak at $E=4.9$ keV, a width $\sigma$=1.2 keV, and an
equivalent width EW=2.4 keV.
(Figure\,\ref{mkn335_broad_fe}). 
Even though this yields an acceptable fit, the line widths and 
equivalent width seem 
to be unusually extreme for a broad line \citep{guainazzi06},
although the presence of such a line was suggested in earlier observations
\citep{gondoin02, longinotti07a}.
This additional component is not necessary when a partial covering model 
is applied to the data.

The narrow 6.4 keV Fe line is present in the high- and low-flux spectra.
The fluxes in the Fe K$\alpha$ 6.4 keV line in the 2006 and 2007 \xmm\ observations
are  (1.83\plm0.33)$\times 10^{-16}$ W m$^{-2}$ and (1.15\plm0.45)$\times 10^{-16}$
W m$^{-2}$, respectively. This result suggests that the flux in the Fe K$\alpha$
6.4 line has become fainter in the low X-ray flux state compared to the high state. 
The equivalent widths ($EW$) however, are
  different due to the much
lower continuum flux during the 2007 observation. We found $EW=100$\plm20 eV and 200\plm80 eV 
during the 2006 and 2007 observations, respectively. 
 While the 6.4 keV Fe K$\alpha$ line is clearly present in both spectra,
the 7.0 keV Fe XXVI Ly$\alpha$ line that was found in the 2006 \xmm\ observation 
\citep[][]{oneill07}
is absent during the low state in 2007. 

\begin{figure}
\epsscale{0.8}
\plotone{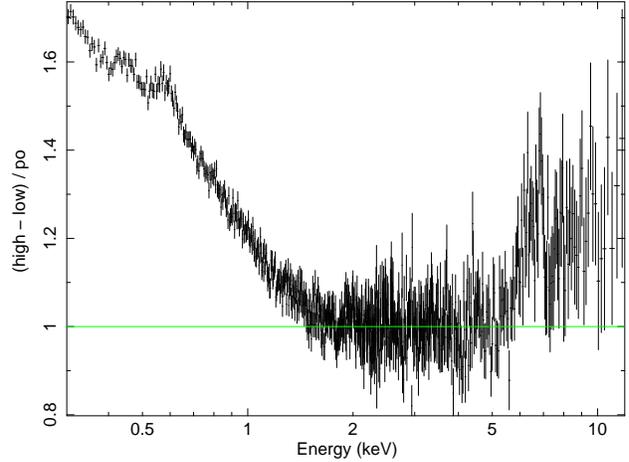}
\caption{\label{mkn335_2006_2007_pc}
Difference spectrum between the 2006 high-state and 2007 low-state \xmm\
spectra fitted in the 2-5 keV range
by a Galactic absorber power law with \ax=1.3 and extrapolated to lower and higher
energies. 
} 
\end{figure}

Figure\,\ref{mkn335_2006_2007_pc} displays the difference spectrum between the 2006 high-state and 2007
low-state fitted in the 2-5 keV energy range with an Galactic absorbed power law (\ax=1.3) and then
extrapolated to lower and higher energies. This difference spectrum clearly shows that the changes
between the high and the low state happened primarily at soft energies below 1.5 keV. In some ways this
plot is similar to the high and low-state spectral energy distributions shown in Figure 2 in
\citet{grupe07b}. Obviously the high and low state spectra cannot be described simply by 
a renormalization of a single power law model. 
Despite the fact that most of the flux change between 2006 and 2007 occurs 
at lower energies, the soft X-ray spectral slopes in the 2006 and 2007 \xmm\ 
observations  are similar. From a broken power law model fit the soft X-ray spectral slopes of the 2006 high
and 2007 low states are \ax=1.73\plm0.01 and \ax=1.89\plm0.04, respectively.
At higher energies, however, the energy spectral indices are significantly different. 
During the 2006 \xmm\ observation the slope was about \ax=1.0, which is quite 
typical for a NLS1 \citep[e.g. ][]{lei99a}. The spectral slope in the 2007 \xmm\ 
observation is very hard with \ax=$-$0.1. We investigate two models that can explain such a 
spectral shape: a partial covering absorber and reflection from an ionized accretion disk.

\subsubsection{The Partial Covering Absorber Scenario}

Partially covering a single power law continuum 
significantly improves the fit to the low-flux 2007 spectrum compared to 
the broken power law model.
The models still requires three Gaussian profiles to fit the
low energy emission features and the narrow 6.4 keV emission line 
($\chi^2/\nu=380/299$).
The broad Gaussian profile 
around 5 keV used in the broken power law model is not required. 
 In this scenario the
intrinsic power law has an energy spectral slope \ax=1.78.  The absorber
has a column density of 
 $N_{\rm H}=15.1\times 10^{22}$ cm$^{-2}$ and a covering fraction $f_c$=0.94.
These partial covering parameters are comparable
to the ones found by \swift\ during the 2007 May and June observations \citep{grupe07b}.
The partial covering parameters for the 2006 observation are then
$N_{\rm H,pc}=5.5\pm0.2\times 10^{22}$ cm$^{-2}$ and $f_c$=0.32\plm0.01.

Even though a comparison of the 2007 \xmm\ data
with the 2007 May and June \swift\ observations suggests some variability 
in the column density, the spectral parameters are consistent during these low-flux periods. Notably, 
all spectra fitted by the partial covering absorber model only require a single underlying power law spectrum.
The spectral slope during all these low-state observations remains fairly constant within uncertainties at \ax=1.8.
The column densities were always within a factor of two.  
The covering fraction at all four epochs was $f_c$=0.94.
The column density was highest during the 2007 May 17 \swift\ observation with $N_{\rm H,pc}=2\times10^{23}$ cm$^{-2}$
and  lowest on May 25 with   $N_{\rm H,pc}=1\times10^{23}$ cm$^{-2}$ \citep{grupe07c}.
The column densities during the
2007 June \swift\ and 2007 July 10 \xmm\ observations are in between those values with $N_{\rm H,pc}=1.5\times10^{23}$ cm$^{-2}$.
The covering fraction during all these four observation was $f_c$=0.94.

One way to test whether the partial covering absorber model is self-consistent,
we fitted the 2006 and 2007 pn data simultaneously in XSPEC. This test can show if the
same underlying continuum spectrum can be applied to the data only affected by the 
absorber column density and covering fraction. 
The first test was to use a 
model with two power laws (replacing the broken power law)
and one partial covering absorber. While the power law
parameters were tied, the partial covering absorber parameters were left free to vary,
first only the covering fraction and second also the absorber column density. In none of
these cases could we obtain an acceptable fit to the spectra. This result suggests that 
high and low states have different underlying continua.
The situation changes when 
adding an additional partial covering absorber to the model (XSPEC model {\it wabs * zpcfabs
* zpcfabs * (powl + powl)}).
This model significantly improved the
fit from a reduced $\chi^2$=2.67 using the one partial covering absorber to
$\chi^2$=1.58 with the two partial covering absorbers.  One of the partial covering
absorber components, however, requires a column density of 3$\times 10^{24}$
cm$^{-2}$. The covering fractions, which were left as a free parameter, of this component
are $f_c$=0.2 and 0.8 for the high and low states, respectively.

As mentioned by \citet{grupe07c}, when correcting the fluxes during the low-state 
observations for Galactic and intrinsic absorption, the 
intrinsic fluxes are similar and the intrinsic long-term X-ray flux amplitude 
varies by a factor of about 4-6 \citep{grupe07c} which 
is similar to other NLS1 galaxies \citep[e.g.][]{grupe01}.
The unabsorbed fluxes in the 0.2-2.0 keV and 2.0-10.0 keV band during the 2006
\xmm\ observation were 6.7 and 6.4 $\times10^{-14}$ W m$^{-2}$, respectively, which
is very similar to the low-state \xmm\ observation on 2007 July 10, which were 6.9
and 5.4 $\times 10^{-14}$ W m$^{-2}$, respectively. Note that during the extreme
low-state on 2007 May 17 observed by \swift, the fluxes were 4.1 and 3.2 $\times
10^{-14}$ W m$^{-2}$, respectively.  
 These values do not suggest significant
intrinsic flux variability. Note that the absorption-corrected
rest-frame 0.2-2.0 keV flux
during the ROSAT All-Sky Survey \citep[RASS;][]{voges99} 
was 4$\times 10^{-14}$ W m$^{-2}$ (Table\,\ref{lc_summary}), 
very similar to the flux
found during the 2007 May 17 \swift\ observations. 

\subsubsection{A reflection interpretation for \mrk}

The 2007 low-flux, broadband X-ray spectrum of Mrk 335 can be well fitted with a single
power law continuum and blurred ionized reflection ($\chi^{2}_{\nu}=1.1$).
It has been shown that in some NLS1s the same low-flux model can be applied
to the high-flux state by simply re-normalizing the power law component
(e.g. Fabian et al. 2004; Gallo et al. 2007a).  However, this is not the
case for Mrk 335 as re-normalizing the power law results in a poor fit to the
2006 high-flux state.  Allowing both the power law and reflection components
to vary freely cannot simultaneously describe the high-energy ($E=2-10$ keV)
and low-energy ($E<2$ keV) spectra of Mrk 335 during the high-flux state.
Similar difficulties were encountered by O'Neill et al. (2007) with the same data set
and demonstrate the need for an additional continuum component during the
high state.

One possibility proposed to describe the flaring in I~Zw~1 \citep{gallo07a, gallo07b} 
was that different parts of the accretion disk are
illuminated by different continuum spectra.
To model this scenario we introduce a second power law with (Model A) and
without (Model B) associated reflection.
In Model A we assume that a
steep power law (with photon index $\Gamma_1$) illuminates the inner
accretion disc from $r_{\rm{in}}$ to $r_{\rm br}$, while a flatter power
law ($\Gamma_2$) is associated with outer disc reflection from
$r_{\rm{br}}$ to $r_{\rm out}$. In Model B, no reflection is associated
with the steep power law, which may be explained if that component is
beamed away from the disc; on the other hand, in Model B we assume a
broken power law emissivity profile for the single reflector with an
inner emissivity index $q_{\rm in}$, breaking to $q_{\rm out}$ at
$r_{\rm br}$. The two different continuum slopes may in fact be due to
different mechanisms (e.g. bulk motion Comptonization and standard
thermal Comptonization, see e.g. Niedzwiecki \& Zdziarski 2006) 

In addition, a MEKAL hot diffuse gas component with Fe abundance tied
to that of the reflector(s) is included to account for any distant
emission due to thermal plasma which should account for the ionized Fe
L lines tentatively detected in the RGS during the lower-flux
state. The parameters of this component are linked in all
observations. As with any other model discussed above, we also include
a narrow Fe K$\alpha$ emission line at $\sim$~6.4~keV and a $\sim
7$~keV line, which is however significantly detected only in the high
flux state 2006 \xmm\ observation.
A narrow Ne line with constant intensity at
$\sim 0.9$~keV is also included in all fits.

Both Model A and B describe the various flux states well
(Figure\,\ref{mkn335_2006_2007_ref}).
See Tab.~\ref{reflection_model} for fit
parameters.  In both models the difference between the high- and
low-flux state is primarily attributed to the significantly diminished
flux of the steeper power law component.  While the two models differ
in their interpretation of the second power law (e.g. beamed versus
un-beamed continuum emission), these are issues that can be tested with
further analysis (Gallo \et in prep.). Both models are successful in
describing the spectra, but they both require additional components to
explain the soft excess. 

\subsubsection{Hard X-ray BAT spectrum}
Mrk 335 is detected in the 22 month \swift\ Burst Alert Telescope
\citep[BAT;][]{barthelmy05} 
 AGN survey \citep{tueller08}. The BAT spectrum can be fitted by
a single power law with \ax=1.77$^{+0.87}_{-0.64}$. This is a similar slope as
found in the \swift\ XRT and \xmm\ low-state data fitted with a single power law
and partial covering absorber.  Figure\,\ref{mkn335_xmm_bat} displays the \xmm\
pn and \swift\ BAT spectra fitted simultaneously in {\it XSPEC} with a power law
with a partial covering absorber with \ax=1.85\plm0.03, $N_{\rm H,
pc}=(16.4\pm1.0)\times 10^{22}$ cm$^{-2}$, and a covering fraction of
$f_c$=0.94\plm0.01 ($\chi^2/\nu$=582/315). The normalizations of the power law
models were left free. Note that the BAT spectrum is the
average spectrum over 22 months, so it is not simultaneous with the
\xmm\ low state spectrum. The observed 20-40 keV and 20-100 keV fluxes are
7.6$\times 10^{-15}$ W m$^{-2}$ and 1.33$\times 10^{-14}$ W m$^{-2}$, which is
in agreement with the measurements from the 2006 June {\it Suzaku} observation
\citep{larsson07}. We do not know from the BAT data if the flux in the
high-energy bands was lower during the low-state in 2007 May to July. However, 
the AGN in the 22 month survey do not show strong variability compared with what
is typically observed at lower energies (J. Tueller, priv comm, 2008).

\subsubsection{Soft X-ray emission lines in the RGS \label{rgs}}

The X-ray slopes obtained from single power law fits to the
EPIC pn (0.2-2.0 keV) and RGS (0.35-2.2 keV) spectra are consistent
(\ax=1.7) for the 2006 and 2007 observations.

The low-flux, broad-band X-ray spectra of Mrk 335 in Figure\,\ref{mkn335_xray_epic}
shows significant residuals below 2 keV when fitted with a single power law. 
The moderate-resolution spectrum suggests
the presence of an ionized gas either in absorption or emission, or both. 
Figure\,\ref{mkn335_rgs} shows the higher resolution RGS spectrum of Mrk 335 fitted 
with a single power law model. The
residuals from this fit clearly show the presence of strong emission lines around
13, 19, 22, and 25 \AA. 

The combined RGS 1 and RGS 2 spectrum is shown in the lower four panels of
Figure\,\ref{mkn335_rgs} in the wavelengths bands:
7-14\AA, 13-21\AA, 20-28\AA, and 27-34\AA.  Possible identification of the 
most likely features are reported.
Most of these lines seem to be highly ionized Fe lines including some Fe group
elements like Mn or Ni. We could also identify Ne IX at 13 \AA. This line,
however, is not as strong and prominent as was found in  the spectrum of the partial covering
candidate and highly optically polarized NLS1 Mrk1239 \citep{grupe04c}. 
The most prominent group of lines that we have identified is the Helium K$\alpha$ 
complex of O VII emission lines around 22 \AA. These lines can potentially be
used for line diagnostics to measure properties of the line emitting plasma (\S 4.2).

\subsection{UV/Optical Data}

\subsubsection{UVOT and OM photometry}

The UV/optical photometric measurements of Mrk 335 made with the
\xmm\ OM and \swift\ UVOT are listed in Table\,\ref{mkn335_uv}.
The values given in the table are corrected for Galactic
reddening \citep[$E_{\rm B-V}$=0.035; ][]{sfd98} and the OM magnitudes were
adjusted, as described in \S 2, to be comparable with the \swift\ UVOT magnitudes.
The data suggest that Mrk 335
was slightly brighter in the Optical/UV in 2000 than in 2007.
However, the June 2007 \swift\ and \xmm\ measurements suggest some rebrightening.
The origin of the relatively large
deviation in the UV W1 magnitude during the 2007 \xmm\ observation is not
obvious. The 5 reference stars listed in Table\,\ref{ref_stars} show a large
scatter between the \xmm\ and \swift\ observations ranging between 0.47 to 0.96
mag. However, they all suggest that the UV W1 magnitudes between the OM and UVOT
have to be adjusted significantly. 

As mentioned in \citet{grupe07b}, the optical to X-ray spectral slope \aox
\footnote{The
X-ray loudness is defined by \citet{tananbaum79} as \aox=--0.384
log($f_{\rm 2keV}/f_{2500\AA}$).} of Mrk355 was between 1.9 and 1.65 
during the X-ray flux low-state. According o the relation given in \citet{strateva05} a \aox=1.3 is expected,
which is exactly what has been seen during the high-state \citep{grupe07b, gallo06}.

\subsubsection{Optical Spectroscopy}

The upper panel of 
Figure\,\ref{mkn335_fe} displays the optical spectrum of Mrk 335 taken on 
2007 September 07 at the 
Xinglong Observatory. The spectrum appears nearly identical compared to the spectrum taken 8 years earlier
at the 1.52m telescope at ESO La Silla
\citep{grupe04}. The continuum slope and emission line fluxes 
are consistent within the errors. 
In Table\,\ref{line_summary} line measurements are given for features between
the He II4686 and Fe X6375 emission lines. The line measurements in the H$\beta$ and
[OIII]5007 lines are, within the errors, the same as given by \citet{grupe04}.

Mrk 335 displays strong coronal Fe lines of [Fe VII]5159, [Fe XIV]5303,
[Fe VII]5721,
6087 and [Fe X]6375 (Figure\,\ref{mkn335_fe}). 
  Even though coronal Fe line are found often in
 Seyfert 1 galaxies \citep[e.g. ][]{erkens97,mullaney07}, 
 seeing these optical lines 
so strong, as in Mrk 335, is rather rare. 
 Similarly 
strong coronal Fe line has been found in the NLS1 galaxies Mrk1239 and 
Ark 564 \citep{erkens97}, 
and in the Seyfert 1.5 MCG-06-15-30 \citep{reynolds97}.
Unusual for a NLS1, Mrk 335 also exhibits  strong HeII4686 and HeI5876 lines
in the present spectrum, which was also detected previously
(e.g., Kassebaum et al. 1997). Typically, this feature is faint in
NLS1 galaxies \citep[e.g.][]{grupe04}.

\section{\label{discuss} Discussion}

We reported the \xmm\ observations of the NLS1 Mrk 335 obtained in
2007 July 10, confirming the low-state of Mrk 335 
found from \swift\ observations in 2007 May/June \citep{grupe07c}. 
Mrk\,335 is remarkable in the sense of being still bright enough
in its low-state to allow a detailed spectral analysis. 

The X-ray spectra of a substantial fraction of
(NL)S1 galaxies are complex (e.g., Gallo 2006), and different types of models
have been considered to explain them, 
including reflection (e.g., Ballantyne et al. 2001, Crummy et al. 2006),
partial covering (e.g., Tanaka et al. 2005) and relativistically smeared 
ionized absorption
(e.g., Done et al. 2007). Variability induced by (cold) absorption is quite common
in AGN, and most frequently observed in Seyfert 2 galaxies and intermediate type
Seyferts. An extreme example is the galaxy NGC\,1365 which shows
highly variable cold absorption along the line of sight on the time scale
of days (Risaliti et al. 2007).      

Partial covering models have been frequently applied
to the X-ray spectra of AGN and NLS1 galaxies in particular (e.g., 
Grupe et al. 2004b,c, Gallo et al. 2004, Pounds et al. 2004, Tanaka et al. 2005),
and also to Galactic X-ray binaries (e.g., Brandt et al. 1996, Ding et al. 2006).  
However, origin and confinement of the required dense blobs of gas
in AGN partial coverers are not well understood (e.g., Kuncic et al. 1997).
Reflection models generally provide similarly successful X-ray spectral fits.  
Apart from spectral modeling, temporal variability is an 
important discriminant between these two types of models.  
The deep low state of Mrk 335 discovered with \swift\ \citep{grupe07c}
and still present during our follow-up \xmm\ TOO observation, 
enables us to do a detailed comparison of partial covering
and reflection models.  
%

\subsection{X-ray continuum}

Before we compare the models themselves we compare Mrk 335 with
several potentially similar NLS1 galaxies. 
The low-state X-ray spectrum of Mrk 335 looks in many ways like the one seen during the 
low-state in NGC 4051 \citep{guainazzi98}. While \citet{pounds04} argue for 
a partial covering absorber scenario, \citet{ponti06} argue for a
 reflection dominated spectrum.
While a fit to the mean low-state spectrum of NGC 4051 using a 
partial covering or a reflection model 
results statistically in very similar fits, \citet{ponti06} could show that 
a reflection model can reproduce their
findings in the rms spectra and flux-flux plots better than a partial 
covering absorber model. Due to the relatively short length of our \xmm\ ToO
observation the low-state data of Mrk 335 are not of sufficient
quality to do such an analysis. 

Mrk 335 also displays many similarities to the NLS1 galaxy Mrk1239:
a) an usual X-ray continuum spectrum,
b) strong X-ray emission lines \citep{grupe04c},
and c) strong optical Fe coronal lines \citep{erkens97}.
The X-ray spectrum of Mrk1239 was interpreted by \citet{grupe04c} with
a partial covering absorber.
While Mrk1239 is highly polarized with more than 4\% at optical
wavelengths \citep{goodrich89, grupe04c},
Mrk 335 does not show any significant degree of polarization \citep{berriman89,
berriman90, smith02}.

The column density of the partial coverer 
we infer from our spectral fits is similar to those
often observed in broad absorption line (BAL) quasars \citep[e.g. ][]{grupe03}.
In this context, it is interesting to point out the potential
similarity of Mrk 335 with the highly variable NLS1 galaxy WPVS007
which dropped by a factor of more than 400 
between the ROSAT All Sky-Survey observation and follow-up 
X-ray observations \citep{grupe95,grupe07b}.
Neither NLS1 galaxy shows any detectable changes in their optical spectra even though their X-ray
fluxes changed substantially. While Mrk 335 lacks UV spectroscopy during the low state,
the UV spectrum of WPVS 007 showed the emergence of strong BAL absorption features 
\citep{leighly07}. Recent \swift\ observations of WPVS 007 suggest that the X-ray spectrum is
described by a partial covering absorber (Grupe et al. 2008, in prep). 

An interesting connection can also be drawn to the X-ray weak quasar PG 2112+059
\citep{schartel07}. Also here the X-ray spectra can be fitted by partial covering absorber
and reflection models. Although the fits are statistically similar (like what we saw in Mkn
335) the authors prefer the blurred X-ray ionized reflection model, because the absorber
model yields unreasonable high column densities and ionization parameters. 
In general, studying bright AGN in low-state
like Mrk 335 gives us access to well-exposed X-ray spectra of
X-ray weak AGN 
which appear to be significantly fainter in X-rays than what is expected from
optical luminosities \citep{brandt00}. Typically these AGN are too faint in X-rays to
obtain well-exposed X-ray spectra. Note that the situation in Mrk 335 is different compared
with the X-ray weak quasar PHL 1811 which is intrinsically X-ray weak \citep{leighly07a}.

\paragraph{Spectral Fits.} 
The spectral fits to the X-ray spectra of Mrk 335
involving a partial coverer are relatively simple, in the sense
of only involving two free parameters to describe the partial coverer itself.
All low-state spectra of Mrk 335  can be fitted by the same 
underlying single power law with \ax=1.8. 
However, the \xmm\ high state data cannot be fitted 
by the same underlying power law.
The 2006 \xmm\ data
do require an underlying broken power law model \citep{grupe07c}.  
We performed several tests with double power law with one and two partial covering
absorbers. We found that we can fit the high and the low states with the same underlying
continuum model, but the low state spectrum then
requires a very high column density of several
times $10^{24}$ cm$^{-2}$ to describe the spectrum at energies $>$ 2 keV. 
On the other hand,
a constant flux of the underlying continuum may explain why the optical spectrum has not
changed at all between 1999 and 2007.  

The steep X-ray spectral slope \ax=1.8 continues even into the \swift\ BAT hard X-ray energy range.
While X-ray spectral slopes of \ax=1.8 are quite common for NLS1s in the 0.2-2.0 keV range \citep{bol96,
grupe01}, this steep X-ray spectral slope is unusual at hard X-ray energies. In
the BAT survey \citep{tueller08} no
AGN has been found with an X-ray spectral slope as steep as Mrk 335. Note that also in the 2-10 keV range an X-ray
spectral slope \ax=1.8 is quite unusual \citep{brandt97, lei99b}. As suggested by e.g. \citet{pounds95} steep hard
X-ray spectral slopes can be caused by thermal Comptonization of UV photons of the hot accretion disk corona.

\paragraph{X-ray and UV variability.} 
The short-time variability found during the low-state observation
shows that the bulk of the soft X-ray emission is not from an
extended scattering region
as suggested by \citet{grupe04c} for Mrk1239 or \citet{immler03} for Mrk6.
 These regions are typically too large to show
variability on time scales of hours. Note that the mass of the central black hole in
 Mrk 335 is of the order of  $1.5-3\times 10^7$\msun\ \citep{peterson04,vestergaard06},
which is on the high-mass end of the black hole mass distribution
of NLS1 galaxies \citep{grupe04b, watson07}.

The variability detected during the low-state observation 
implies a source size of a few light hours (see Figure\,\ref{mkn335_pn_lc})
which 
in turn requires that the partial coverer has to 
consist of very compact, dense clumps of gas.


Perhaps in favor of the partial covering interpretation is the 
variability in the UV/optical band (even though small). 
In the two bluest filters, W2 and M2, Mrk 335 varies stronger by
0.2 mag than in the optical V and B
filters. Since the UV magnitudes seem to follow the X-ray 
count rates, this might indicate that the variability in the UV and in X-rays
is caused by the same mechanism.  However, longer-term monitoring 
of the UV and X-ray variability is required to put this
statement on a more firm basis. 


\paragraph{Models of partial coverers.}
In summary, the most successful phenomenological partial coverer model 
consists of very compact blobs of gas covering the variable
X-ray source and is mostly located along our line of sight
in order to ensure non-variable BLR line emission and 
in order to be consistent with the absence of a detectable strongly variable
6.4 keV iron line from the partial coverer itself.  
The requirement of the compactness of the clouds, in
combination with their high column density, emphasizes 
well-known challenges to the partial coverer scenario
including questions like:  
what is the origin of the blobs, and how are they confined 
(e.g., Guilbert \& Rees 1988, Celotti et al. 1992, Kuncic et al. 1997) ? 
In the model of Abrassart \& Czerny (2000), fragmentation
of the inner accretion disk leads to the formation 
of compact clouds which can partially obscure the central
X-ray source, confined perhaps by strong
magnetic fields (Kuncic et al. 1997).
These clouds would reprocess some radiation into
the optical and UV band (Celotti et al. 1992,
Kunic et al. 1997, Abrassart \& Czerny 2000). 
If their emission does contribute
to the total UV emission, then the variability in
the X-ray band would be higher than that in the UV band, as observed. 

\paragraph{Reflection models.} 

Blurred reflection models invoking a single continuum source and reflecting medium have
been shown to fit the 2000 high-flux state observation \citep{crummy06} and the
2007 low-flux state observation (Section 3.2.2).   However, there are a number of
discrepancies between parameter values (e.g. inclination and iron abundance) that indicate
these models are not consistent.  Likewise, the high signal-to-noise 2007 spectrum of
Mrk 335 cannot be modeled self-consistently in the high- and low-energy ranges
(O'Neill et al. 2007; Section 3.2.2). 


The fact that the low-flux spectra of \mrk\ cannot be fit with the
same reflection model used for the higher flux states suggests that
the most straightforward application of light bending does not
necessarily apply \citep{miniutti03, miniutti04}.  
That is, the low-flux state is not solely due to a
single continuum source whose intensity appears diminished because of
its proximity to the black hole while the reflection is kept constant.
One possibility is that the entire disc does not ``see'' the same
continuum source.  Such a scenario was proposed by Gallo \et (2007a,b)
to describe the behavior of another NLS1, I~Zw~1.  Here we considered
this possibility for \mrk. 

Two models are suggested, both rendering equally good fits.
Both scenarios incorporate two power law components, each one illuminating different parts of the
accretion disc.  One component could be associated with a jet producing either 
beamed (Model B) emission
with no reflection component, or un-beamed (Model A, see Ghisellini et al. 2004) emission that
preferentially illuminates the inner-most region of the disc. A jet seen in Mrk 335 is
plausible. Mrk 335 has a clear detection of a compact source at 4.8 GHz \citep{leipski06}. 
The second power law
is associated with the traditional idea of the corona, which blankets the disc
to larger radii.
These, of course, are not exclusive interpretations.
The two different continuum slopes may in fact be due to
different mechanisms (e.g. bulk motion Comptonization and standard
thermal Comptonization, see e.g. Niedzwiecki \& Zdziarski 2006).  

Statistically, both models fit the data very well ($\chi^{2}_{\nu}=1.06-1.10$),
but more importantly in a self-consistent manner.  The primary difference between the 
high- and low-flux states is the diminished flux received by the observer from the inner power law
component (i.e. the one perhaps associated with a jet).  


According to our models A and B, there are situations (especially in the low
state) when the X-ray spectrum of Mrk 335 is reflection-dominated. If so,
the observed flux cannot be used in a simple way to infer the intrinsic X-ray
luminosity of the AGN. In general, the power law component contributes to
the observed flux at infinity, and it also illuminates the accretion disc
giving
rise to the associated reflection component. The intrinsic luminosity of that
power law must take into account not only the flux observed at infinity, but
also that needed to produce the reflection component. If $L_{\rm{obs}}$ is the
power law luminosity observed at infinity and $R$ is the reflection fraction
of the associated reflection component, the intrinsic power law luminosity is
$L_{\rm{int}}=L_{\rm{obs}} (R+1)/2$, which, in the isotropic case ($R=1$),
simply reduces to $L_{\rm{obs}}$. Applying the above arguments to
model A, we estimate that the intrinsic luminosity of all power law components
is $\sim 2.3\times 10^{43}$~erg~s$^{-1}$~cm$^{-2}$ in the high-flux state
and $>= 1.9\times 10^{43}$~erg~s$^{-1}$~cm$^{-2}$ in the low-flux state
(all luminosities
are unabsorbed and given in the 2-10 keV band). In other words, the intrinsic
luminosity of the AGN would be similar in the two states despite the dramatic
observed flux variation.

In the case of model B, we have one power law with its reflection (for
which the same arguments as above can be applied) and one power law with no
associated reflection. The latter could be due to a jetted component beaming
the radiation away from the disc (so that there is no reflection). Hence, we
can estimate its intrinsic luminosity by assuming that the radiation is beamed
in the half-sphere away from the disc ($L_{\rm{int}}<=L_{\rm{obs}}/2$).
For model B, considering the contributions of both the jetted and the non-jetted
components, we then obtain intrinsic luminosities in the range
$(6-6.8)\times 10^{43}$~erg~s$^{-1}$~cm$^{-2}$ in the high-flux state and
$\sim 2.4\times 10^{43}$~erg~s$^{-1}$~cm$^{-2}$ in the low-flux state (the
jetted contribution is negligible in the low-flux state).

\paragraph{Partial covering absorber vs. reflection models.} 
Statistically the reflection models yield  better fits compared with the 
partial covering absorber model (reduced $\chi^2$=1.10 vs. 1.27, respectively).
Both,  the partial
covering absorber model and the reflection models require underlying broken power law
or double power law model 
spectra to fit the high-states of Mrk 335.  
For both models the same underlying continua can be used in the high and low states just
modified by either absorption or reflection. The parameters however, used for the partial
covering absorber model then require that the absorption column density of one of the
partial covering absorber components is large on the order of
 several times $10^{24}$ cm$^{-2}$.
Even though this is not impossible, it requires extremely dense compact clouds to pass
through the line of sight.

Reflection models have been used in the past to describe the X-ray spectra of may AGN
\citep[e.g.][]{fabian07} and have also been applied successfully 
to describe the high-state X-ray spectra of Mkn
335 \citep[e.g.][]{ballantyne01, crummy06, longinotti07a, oneill07} and we have shown here
that they also consistently produce the low-state. Further, we mention in passing that we
cannot exclude a third possibility - that the spectrum is dominated in high-state by
reflection, and this reflection spectrum is modified by partial covering absorption. 
A scenario like this has been suggested e.g. by \citet{chevallier06}
and \citet{merloni06}. 
In some ways the situation in Mrk 335 
is similar to that of PG 2112+059 where the absorber model also
yields very high column densities which seems to favor the reflection model. On the other
hand the partial covering absorber model is rather simple and does not require many
parameters to fit the data reasonably well.  At the end one has to conclude that both
models are valid to explain the spectral shape seen in the low state spectrum of Mrk 335.
Nevertheless with the low-state \xmm\ observation presented here we were able to put
strong constrains on both the partial covering and reflection model parameters. In order to
really distinguish between both models either a very long \xmm\ observation in low state is
needed to search for the deviation between both models at intermediate energies between 2-6
keV or it is necessary
to obtain a spectrum at energies $>$ 10 keV using e.g. \suzaku.

\subsection{X-ray emission lines}

Several low-energy ($E<2$ keV) X-ray emission
lines are detected in the RGS spectra of Mrk 335. 
Soft X-ray emission lines are typically found in obscured Seyfert 2 type
AGN \citep[e.g.][]{guainazzi07, kinkhabwala02} and are only seen
in Seyfert 1s when the 
AGN enters a low-flux state.  Consequently, low-energy emission lines have only 
been detected in a few Seyfert 1s like:
Mrk590, NGC 4051, and NGC 5548 \citep[][respectively]{longinotti07c,
pounds04, steenbrugge03}. 
This is the first time such lines have been observed in Mrk 335.
Even though these lines are expected in Seyfert 1s when the continuum is suppressed, as in the low state 2007
\xmm\ observation, they are very prominent even with just a 22 ks RGS exposure.

Most notable in the RGS spectrum of Mrk 335 is the O VII emission line complex around 22\AA. 
The forbidden (f) line at 22.1012 \AA~ 
and the intercombination (i) line at 21.8044 \AA~ are secure identifications. 
These lines have also been found in the low-state RGS spectra of NGC 4051 \citep{pounds04, ponti06}.
The resonance (r) line appears to be weak and slightly red-shifted with respect to its rest 
frame wavelength of 21.602 \AA, which could be due to resonance scattering and absorption. 
Although the line intensities are considerably uncertain, the presence of the O VII line 
complex suggests coronal emission from collisionally ionized plasma. In principle the 
lines from this complex can yield reasonably precise information on the temperature, 
densities, and ionization state of the source. However, owing to the uncertainties 
we conclude that it is 
not useful at this time to attempt a detailed line ratios analysis such as the G = (i+f)/r 
ratio or the R = f/i ratio \citep[e.g. ][]{pradhan81}.
Nor is it possible to deduce the existence of additional features widely seen in warm absorber 
spectra \citep[e.g.,][]{lee01},
such as the OVI KLL absorption features at (22.05,21.87) \AA, 
lying between the f and i lines \citep{pradhan00, pradhan03}.
In addition to O VII, we have identified, to varying degree of certainty, a number of iron 
lines from coronal ions such as Fe XVII to Fe XXV. The iron lines complement the independent 
observations of the UV/Optical lines from coronal iron ions in lower ionization stages 
discussed in the next section. Note, that the RGS data have independently been analyzed by
\citet{longinotti08} who  present line ratios. They did not find a unique solution of the 
plasma properties either, but concluded that the location of the X-ray line emitting gas 
is within 0.01 pc.

\subsection{Optical emission lines}

The optical emission-line spectrum of Mrk 335 is remarkably
constant between the observations taken 8 years ago, and the one
taken nearly simultaneously with the X-ray low-state observations.
Since the broad component of the Balmer lines and the broad
component of the HeII line are very sensitive to the EUV-to-soft X-ray part
of the continuum, the observed lack of variability implies
that the bulk of the BLR still sees the high-state continuum.
In the context of the partial covering model, this can be understood if
the absorber has a small global covering fraction and is mostly located
along our line-of-sight.

The highly ionized coronal Fe lines found in the optical spectrum of Mrk 335 are
rather strong. While coronal Fe lines are present in the 
optical spectra of Seyfert 1s, they typically appear faint \citep[e.g.][]{dan01}
with typical flux ratios to [OIII]$\lambda$5007\AA\ about 0.03 to 0.06. 
In comparison with 
the AGN sample of \citet{erkens97}, only 2 out of 15 AGN (Mrk486 and Akn 564), have 
coronal Fe lines stronger than Mrk 335. 
The central wavelengths of the 
iron lines in Mrk 335 suggests they
are blueshifted with respect to the [OIII] lines, as is commonly seen in
coronal lines (e.g., Erkens et al. 1997, Rodriguez-Ardila et al. 2006). 
The 5\AA\ resolution of our spectrum, however, does not allow a
precise measurement of the central wavelength of the lines and therefore 
the shift of the line with respect to the rest-frame cannot be
determined with high accuracy. 
As generally observed \citep{erkens97}, the line widths
of the coronal Fe lines are between the NLR 
(FWHM$\approx500$ km s$^{-1}$) and the BLR (FWHM(H$\beta$)=1700 km s$^{-1}$).

\section{Conclusions}

We report the \xmm\ and \swift\ observations of the bright NLS1 galaxy Mrk 335 in a 
deep, and unprecedented, low-flux state.  The main findings are as follow:

\begin{itemize}
\item The \xmm\ ToO observation of 2007 July 10 still found Mrk 335 in a
 very low-flux state as previously detected by \swift. With this \xmm\ observation,
  however,
 the parameters of the X-ray spectral models can be well-constrained. Low-state
 observations of Seyfert 1s are rare and Mrk 335 is exceptional in being still bright in
 its low-state. Low states give us access to X-ray emission lines that otherwise cannot be
 seen (see below). The low state of Mrk 335 lasted at least two months. 

\item While a simple broken power law model requires a broad red-shifted Fe Gaussian
 line component with an extremely large equivalent width EW=2.4 keV,
 this line component is not required either in the partial
 covering absorber nor the reflection models.  
The partial covering model can be fit to
high-state and low-state data with the
same underlying double power law model continuum 
when two absorber components are used,
even though formally a second powerlaw
component is not {\em required} to fit
the low-state spectrum.
The reflection model formally accounts
for the two-component nature of the
observed spectrum by multiple reflection
where different parts of the accretion
disk see different continua.

\item High-resolution spectroscopy clearly reveals the presence of 
X-ray emission lines in the 
low-energy X-ray spectrum of Mrk 335, including transitions of oxygen and iron.
These lines are only visible in the low-state.
Preliminary results suggest that the emission-line gas
is collisionally ionized. However, to perform proper
line diagnostics of the ionized gas to derive gas density
and  temperature requires 
a longer RGS exposure of Mrk 335 in its low-state.

\item The optical spectrum of Mrk 335 taken during the X-ray low-state
did not change in comparison to a spectrum taken 8 years earlier.
The constancy of the broad Balmer lines indicates that the bulk of the
broad line region was still illuminated by the ``high-state'' continuum.
The optical spectrum of Mrk 335 displays a variety of strong,
highly-ionized coronal Fe lines. These might be linked to the
component emitting the X-ray lines detected with RGS. The non-variability
 seen in these lines 
may indicate that the intrinsic SED has not changed.
\end{itemize}

Mrk 335 will continue to be monitored with \swift\ in order to trigger another, much
deeper X-ray observation if the AGN enters another low state.

\acknowledgments
This work is based on observations obtained with \xmm, an ESA
 science mission with  instruments and contributions
 directly funded by ESA Member States and NASA. 
In Germany, the XMM-Newton project is supported by
the Bundesministerium f{\"u}r Wirtschaft und Technologie/Deutsches
Zentrum fur Luft und Raumfahrt, and the Max-Planck Society.
We thank the \xmm\ Project Scientist  Norbert Schartel 
for approving our ToO 
request and Michel Breitfellner for preparing and scheduling the \xmm\ observation.
We are very grateful to Ron Remillard for preparing the RXTE ASM light curve,
the \swift\ PI Neil Gehrels for approving the monitoring campaign of Mrk 335
with \swift, and Jack Tueller for providing the 22 month BAT survey spectrum of
Mrk 335.
We thank the Xinglong observatory,
NAOC for providing telescope time, and the support astronomer Jing Wang and his
night assistant Jianjun Jia for performing the optical spectroscopy observations
on 2007 September 07. We also want to thank the anonymous referee for his/her 
helpful and constructive report that improved this paper. 
This research has made use of the NASA/IPAC Extragalactic
Database (NED) which is operated by the Jet Propulsion Laboratory,
Caltech, under contract with the National Aeronautics and Space
Administration, and the Atomic Spectra Database (version 3.1.2)
of the National Institute of Standards and
Technology (physics.nist.gov/asd3). DX acknowledges
the support of the Chinese National Science Foundation (NSFC)
under grant NSFC-10503005.
Swift is supported at PSU by NASA contract NAS5-00136.
This research was supported by NASA contracts NNX07AH67G and NNX07AV04G (D.G.).

\clearpage

\begin{deluxetable}{lcccr}
\tabletypesize{\tiny}
\tablecaption{Summary of the 2007 July 10 \xmm\ observations of Mrk 335
\label{obs_log}}
\tablewidth{0pt}
\tablehead{
\colhead{Instrument} & \colhead{T-start\tablenotemark{1}} & 
\colhead{T-stop\tablenotemark{1}} &
\colhead{$\rm T_{exp}$\tablenotemark{2}} 
} 
\startdata
EPIC pn     & 17:09 & 21:57 & 16779 \\
EPIC MOS-1  & 16:14 & 21:56 & 20348 \\
EPIC MOS-2  & 16:14 & 21:56 & 20351 \\
RGS-1       & 15:50 & 21:58 & 21932 \\
RGS-2       & 15:50 & 21:58 & 21936 \\
OM U        & 15:36 & 16:32 &  3000 \\
OM B        & 16:32 & 17:10 &  2000 \\
OM UVW1     & 17:10 & 18:22 &  4000 \\
OM UVM2     & 18:22 & 19:34 &  4000 \\
OM UVW2     & 19:34 & 20:46 &  4000 \\
OM UVW2     & 20:46 & 21:57 &  4000 \\
\enddata

\tablenotetext{1}{Start and End times are given in UT}
\tablenotetext{2}{Observing time given in s}
\end{deluxetable}

\begin{deluxetable}{ccccccccc}
\tabletypesize{\tiny}
\tablecaption{List of reference stars used to determine the error in the
photometry of Mrk 335 and the off-set between the \swift-UVOT and \xmm-OM filters.
The magnitudes are not corrected for Galactic reddening.
\label{ref_stars}}
\tablewidth{0pt}
\tablehead{
\colhead{Object} & \colhead{$\alpha_{2000}$} & 
\colhead{$\delta_{2000}$} & 
\colhead{Filter} & \colhead{XMM 2000} & \colhead{XMM 2007}
& \colhead{\swift\ 001\tablenotemark{1}} & \colhead{\swift\ 002\tablenotemark{1}}
& \colhead{\swift\ 003-005\tablenotemark{1}} 
} 
\startdata
Star 1 & 00 06 22.8 & +20 14 58.0 & V     & 14.28\plm0.01 & \nodata       &  14.29\plm0.01 & 14.28\plm0.01 & 14.28\plm0.01 \\
       &            &             & B     & 14.94\plm0.01 & 14.90\plm0.01 &  14.89\plm0.01 & 14.87\plm0.01 & 14.86\plm0.01 \\
       &            &             & U     & 14.88\plm0.01 & 14.88\plm0.01 &  14.87\plm0.01 & 14.87\plm0.01 & 14.86\plm0.01 \\
       &            &             & UV W1 & \nodata       & 15.44\plm0.01 &  16.27\plm0.02 & 16.32\plm0.02 & 16.33\plm0.02 \\
       &            &             & UV M2 & 17.43\plm0.16 & 17.45\plm0.09 &  17.90\plm0.06 & 17.91\plm0.05 & 17.94\plm0.05 \\
       &            &             & UV W2 & \nodata       & 17.68\plm0.20 &  17.73\plm0.04 & 17.78\plm0.03 & 17.81\plm0.03 \\
Star 2 & 00 06 32.0 & +20 13 08.3 & V     & 15.46\plm0.02 & \nodata       &  15.48\plm0.02 & 15.48\plm0.02 & 15.51\plm0.01 \\
       &            &             & B     & 16.31\plm0.01 & 16.29\plm0.01 &  16.21\plm0.02 & 16.20\plm0.01 & 16.21\plm0.01 \\
       &            &             & U     & 16.53\plm0.02 & 16.54\plm0.02 &  16.55\plm0.03 & 16.57\plm0.02 & 16.55\plm0.02 \\
       &            &             & UV W1 & \nodata       & 17.26\plm0.03 &  18.30\plm0.07 & 18.20\plm0.05 & 18.17\plm0.05 \\
       &            &             & UV M2 & ---           & ---           &  $>$20.03      & $>$20.35      & $>$20.42      \\
       &            &             & UV W2 & \nodata       & ---           &  19.76\plm0.12 & 19.76\plm0.10 & 19.91\plm0.10 \\ 	
Star 3 & 00 06 17.1 & +20 14 08.4 & V     & 15.40\plm0.02 & \nodata       &  15.46\plm0.02 & 15.43\plm0.02 & 15.42\plm0.01 \\
       &            &             & B     & 16.21\plm0.01 & 16.19\plm0.01 &  16.13\plm0.02 & 16.13\plm0.01 & 16.09\plm0.01 \\
       &            &             & U     & 16.27\plm0.02 & 16.26\plm0.02 &  16.26\plm0.02 & 16.26\plm0.02 & 16.23\plm0.02 \\
       &            &             & UV W1 & \nodata       & 16.88\plm0.03 &  17.69\plm0.05 & 17.83\plm0.04 & 17.74\plm0.04 \\
       &            &             & UV M2 & ---           & ---           &  19.85\plm0.19 & 19.82\plm0.15 & 19.69\plm0.13 \\
       &            &             & UV W2 & \nodata       & ---           &  19.27\plm0.09 & 19.34\plm0.07 & 19.45\plm0.07 \\
Star 4 & 00 06 18.0 & +20 13 17.2 & V     & 15.04\plm0.01 & \nodata       &  15.04\plm0.02 & 15.07\plm0.01 & 15.07\plm0.01 \\
       &            &             & B     & 15.51\plm0.01 & 15.49\plm0.02 &  15.46\plm0.01 & 15.45\plm0.01 & 15.44\plm0.01 \\    
       &            &             & U     & 15.35\plm0.01 & 15.36\plm0.01 &  15.33\plm0.01 & 15.35\plm0.01 & 15.34\plm0.01 \\
       &            &             & UV W1 & \nodata       & 15.70\plm0.02 &  16.35\plm0.02 & 16.34\plm0.02 & 16.37\plm0.02 \\
       &            &             & UV M2 & 16.85\plm0.12 & 16.77\plm0.06 &  16.99\plm0.04 & 17.02\plm0.03 & 17.09\plm0.03 \\
       &            &             & UV W2 & \nodata       & 17.16\plm0.13 &  17.35\plm0.03 & 17.37\plm0.02 & 17.42\plm0.02 \\
Star 5 & 00 06 20.1 & +20 10 50.2 & V     & 14.30\plm0.01 & \nodata       &  14.31\plm0.01 & 14.33\plm0.01 & 14.30\plm0.01 \\
       &            &             & B     & 15.51\plm0.01 & 15.47\plm0.01 &  15.38\plm0.01 & 15.39\plm0.01 & 15.39\plm0.01 \\
       &            &             & U     & 16.45\plm0.02 & 16.45\plm0.02 &  16.46\plm0.03 & 16.41\plm0.02 & 16.48\plm0.02 \\
       &            &             & UV W1 & \nodata       & 17.54\plm0.04 &  18.01\plm0.06 & 18.12\plm0.05 & 18.12\plm0.04 \\
       &            &             & UV M2 & ---           & ---           &  $>$20.03      & $>$20.34      & $>$20.39      \\
       &            &             & UV W2 & \nodata       & ---           &  19.22\plm0.09 & 19.44\plm0.08 & 19.52\plm0.07
\enddata

\tablenotetext{1}{The observing times of the 2007 \swift\ observations are
 listed in Table 1 in \citet{grupe07c}.}

\end{deluxetable}

\begin{deluxetable}{llccl}
\tabletypesize{\tiny}
\tablecaption{Summary of the 0.2-2.0 keV and 2.0-10.0 keV fluxes 
\label{lc_summary}}
\tablewidth{0pt}
\tablehead{
\colhead{Mission} & \colhead{Time\tablenotemark{1}} & 
\colhead{0.2-2.0 keV flux\tablenotemark{2}} &
\colhead{2.0-10.0 keV flux\tablenotemark{2}}  &
\colhead{References\tablenotemark{3}}
} 
\startdata
UHURU      & 1971/1972  & --- & 2.04\plm0.03 & (1) \\
HEAO 1     & 1978-07-07 & --- & 1.1\plm0.2  & (1)  \\
EINSTEIN   & 1981-01    & --- & 1.46\plm0.10 & (1) \\
EXOSAT     & 1983-11-05 & 1.4\plm0.2 & 0.23\plm0,05 & (2) \\
EXOSAT     & 1984-12-06 & 8.3\plm0.1 & 1.54\plm0.03 & (2) \\
GINGA      & 1987-12-03 & --- &  0.91\plm0.16 & (3) \\
GINGA      & 1988-11-29 & --- &  2.00\plm0.10 & (3) \\
ROSAT RASS & 1990-07-14 & 4.00\plm0.10 & ---  & (4) \\
BBXRT      & 1990-12-05 & --- & 0.84\plm0.05 & (1) \\
ROSAT PSPC & 1991-06-30 & 3.40\plm0.10 & --- &  (4) \\
ASCA       & 1993-12-09 & --- &  0.94\plm0.10 & (5) \\
\xmm        & 2000-12-25 & 7.90\plm0.01 &  1.58\plm0.02 & (6) \\
\xmm        & 2006-01-03 & 6.76\plm0.01 &  1.79\plm0.01 & (6) \\
Suzaku     & 2006-06-20 & --- & 1.43\plm0.02 & (7) \\
\swift\ 001  & 2007-05-17 & 0.23\plm0.03 &  0.28\plm0.04 & (6) \\
\xmm        & 2007-07-10 & 0.41\plm0.01 & 0.31\plm0.01  & (8) \\
\enddata

\tablenotetext{1}{Observing time in years}
\tablenotetext{2}{The  0.2-2.0 and 2.0-10.0 keV fluxes are corrected for Galactic absorption and are
give in units of
$10^{-14}$ W m$^{-2}$ (= $10^{-11}$ ergs s$^{-1}$ cm$^{-2}$).}
\tablenotetext{3}{References: (1) \citet{turner93}, (2) \citet{pounds87}, (3)
\citet{nandra94}, (4) \citet{grupe01}, (5) \citet{george00, lei99a}, (6)
\citet{grupe07c}, (7) \citet{larsson07}, (8) this paper
}
\end{deluxetable}

\begin{deluxetable}{cccccccc}
\tabletypesize{\tiny}
\tablecaption{Optical and UV
photometry of Mrk 335 using the  \swift-UVOT and \xmm-OM\tablenotemark{1}
\label{mkn335_uv}}
\tablewidth{0pt}
\tablehead{
\colhead{Filter} & \colhead{XMM 2000} & \colhead{XMM 2007}
& \colhead{\swift\ 001} & \colhead{\swift\ 002}
& \colhead{\swift\ 003} 
& \colhead{\swift\ 004}
& \colhead{\swift\ 005}
} 
\startdata
 V     & 14.00\plm0.01 & \nodata       &  14.22\plm0.01 & 14.19\plm0.01 & 14.19\plm0.01 & 14.15\plm0.01 & 14.18\plm0.01 \\
 B     & 14.15\plm0.01 & 14.25\plm0.01 &  14.49\plm0.01 & 14.42\plm0.01 & 14.44\plm0.01 & 14.41\plm0.01 & 14.42\plm0.01 \\
 U     & 13.00\plm0.01 & 13.17\plm0.01 &  13.29\plm0.01 & 13.20\plm0.01 & 13.23\plm0.01 & 13.22\plm0.01 & 13.23\plm0.01 \\
 UV W1 & \nodata       & 13.52\plm0.01 &  13.20\plm0.01 & 13.08\plm0.02 & 13.09\plm0.02 & 13.10\plm0.01 & 13.05\plm0.01 \\
 UV M2 & 12.92\plm0.16 & 12.90\plm0.01 &  13.12\plm0.01 & 12.95\plm0.01 & 12.98\plm0.01 & 12.97\plm0.01 & 12.93\plm0.01 \\
 UV W2 & \nodata       & 13.07\plm0.20 &  13.19\plm0.01 & 13.01\plm0.01 & 13.04\plm0.01 & 13.04\plm0.01 & 12.98\plm0.01 \\
\enddata

\tablenotetext{1}{All magnitudes are corrected for Galactic reddening \citep[$E_{\rm
B-V}$=0.035; ][]{sfd98} and the \xmm\ OM magnitudes are adjusted as described in
the text to be comparable with the \swift\ UVOT data.
}

\end{deluxetable}

\begin{deluxetable}{llcccccccccccc}
\rotate
\tabletypesize{\tiny}
\tablecaption{Spectral analysis of the 2007 and 2006 \xmm\ spectra of Mrk 335 in the 0.3-12.0 keV range  
\label{xray_res}}
\tablewidth{0pt}
\tablehead{
\colhead{Mission} &
\colhead{Detector} &
\colhead{Model\tablenotemark{1}} &
\colhead{$\alpha_{\rm X,soft}$} 
& \colhead{$E_{\rm break}$\tablenotemark{2}}
& \colhead{$\alpha_{\rm X,hard}$}
& \colhead{$N_{\rm H, pcf}$\tablenotemark{3}}
& \colhead{$\rm f_{c}$\tablenotemark{4}}
& \colhead{$E_{\rm line}$\tablenotemark{5}}
& \colhead{$EW_{\rm line}$\tablenotemark{6}}
& \colhead{$\chi^2/\nu$}
} 
\startdata

\xmm\ 2007 & EPIC pn  &  (a) & 1.63\plm0.02 & --- & --- &  --- & ---  & --- & --- & 3158/312 \\
         &          &  (b) & 1.90\plm0.02 & 1.73\plm0.03 & --0.13\plm0.03 & --- &  --- &
	 --- & ---  & 712/308 \\
	 &          &  (c) & 1.87$^{+0.03}_{-0.06}$ & 1.68\plm0.04 & --0.10\plm0.04 & --- & --- & 6.37\plm0.05 & 0.31\plm0.13 & 505/299  \\
	 &          &  (d) & 1.87$^{+0.03}_{-0.06}$ & 1.77$^{+0.16}_{-0.08}$ &  +0.05$^{+0.09}_{-0.07}$ & --- & --- & 6.42\plm0.04 &
	 0.19\plm0.08 & 333/296
	 \\
         &          &  (e) & 1.78$^{+0.02}_{-0.03}$ & --- & --- & 15.1$^{+1.0}_{-0.9}$ & 0.94\plm0.04 & 
	 6.41\plm0.04  & 0.19\plm0.08  & 380/299 \\
         & EPIC pn + MOS & (a) & 1.55\plm0.02 & --- & --- & --- & --- & --- & --- & 5678/549 \\
	 &               & (b) & 1.89\plm0.02 & 1.73\plm0.02 & --0.19\plm0.03 & --- & --- & --- & --- & 1208/547 \\
	 &               & (e) & 1.78$^{+0.03}_{-0.02}$ & --- & --- & 14.6$^{+1.0}_{-0.6}$  
	 & 0.93$^{+0.04}_{-0.05}$ & 6.41\plm0.03 & 0.26\plm0.07 & 788/538 \\
         & RGS1+2   &  (a) & 1.64\plm0.14 & --- & --- & --- & --- & --- & --- &  424/371 \\
\xmm\ 2006 & EPIC pn  & (a) & 1.62\plm0.01  & --- & --- & --- & --- & --- & --- & 16770/697 \\
         &          & (b) & 1.72\plm0.01 & 1.84\plm0.02 & 1.07\plm0.01 & --- & --- & --- & --- & 1717/695 \\
	 &          & (c) & 1.72\plm0.01 & 1.79\plm0.03 & 1.10\plm0.02 & --- & --- & 6.41\plm0.04 & 0.10\plm0.02 & 1043/686 \\
	 &          & (e) & 1.73\plm0.01 & --- & --- & 5.1\plm0.4 & 0.35\plm0.02 & 6.41\plm0.03 & 0.09\plm0.02 & 834/687 \\ 
         & RGS1+2   & (a) & 1.64\plm0.01 & --- & --- & --- & --- & ---  & --- & 5288/4494 \\
\xmm\ 2006 + 2007 & EPIC pn & (a) & 1.62\plm0.01 & --- & --- & --- & --- & --- & --- & 19930/1008 \\
           &               & (b) & 1.72\plm0.01 & 1.84\plm0.02 & 1.06\plm0.01 & --- & --- & --- & --- & 4131/1006 \\
	   &		  & (c) & 1.72\plm0.01 & 1.67\plm0.03 & 1.28\plm0.02 & --- & --- & 6.41 (fix) & 0.08/0.16 & 1710/997 \\
	   &		  & (e) & 1.70\plm0.01 & --- & --- & 10.8\tablenotemark{7} & 0.56/0.91\tablenotemark{7} & 6.41 (fix) &
			  0.10/0.30\tablenotemark{7} & 2642/997	  \\
	   &		  & (e) & 1.69\plm0.01 & --- & --- & 10.5/14.4\tablenotemark{7} & 0.56/0.93\tablenotemark{7} & 6.41 (fix) &
			  0.10/0.20\tablenotemark{7} & 2575/996	 
\enddata

\tablenotetext{1}{Spectral models used are: 
(a) absorbed power law, 
(b) absorbed broken power law, 
(c) absorbed broken power law with two Gaussians to represent the soft X-ray emission lines below 1 keV and one Gaussian for the 6.4 keV Fe line,
(d) same as (c) but with an additional broad line at 4.9 keV with a $\sigma$=1.5 keV,
 and (e) power law with a partial covering absorber with two soft X-ray Gaussians and one for the 6.4 Fe line.
For all models the absorption column density of the z=0 absorber was fixed to the Galactic value
\citep[3.96$\times 10^{20}$ cm$^{-2}$][]{dic90}.  }
\tablenotetext{2}{The break energy $E_{\rm break}$ is given in units of keV.}
\tablenotetext{3}{Absorption column density of the partial covering absorber $N_{\rm H, pcf}$ in units of $10^{22}$ cm$^{-2}$}
\tablenotetext{4}{Covering fraction $\rm f_{c}$}
\tablenotetext{5}{The line energy $E_{\rm line}$ is given in units of keV.}
\tablenotetext{6}{The line equivalent width is given in units of keV.}
\tablenotetext{7}{Simultaneous fit with a power law with partial covering absorber and Gaussian lines. While the X-ray spectral slope was tied and
the line energies and widths were fixed, the column densities and covering fractions of the partial covering absorber and the normalizations of the
power law continua and the line fluxes were left as free parameters. The values of the column densities, covering fractions, and equivalent widths of
the 6.4 keV line are given for the 2006 and 2007 data. }
\end{deluxetable}

\clearpage

%
%

\clearpage

\begin{deluxetable}{lcccl}
\tablecaption{\label{reflection_model}
Summary of the most important parameters\tablenotemark{a} for the
  best--fitting reflection models (Model A and B; see text for details) 
  to the high-flux (HF) and low-flux (LF)
  state observations. 
  }
\tablewidth{0pt}
\tablehead{
  \colhead{Param.} & \colhead{HF state} & \colhead{LF state} &
  \colhead{tied\tablenotemark{b}} & \colhead{Units}
  }
\startdata
  \hline
  & Model A & ($\chi2_r = 1.06$)\\      
  \hline
  $\alpha_1$   & $1.86^{+0.04}_{-0.02}$ & $1.86^{+0.04}_{-0.02}$ & x & \\    
  $F_{\alpha_1}$  &  $(5.86\pm0.05)\times 10^{-11}$ & $< 3.5\times10^{-13}$ & & [$\times 10^{-3}$W m$^{-2}$]=[ergs s$^{-1}$ cm$^{-2}$] \\    
  $\alpha_2$ &  $0.70\pm0.01$ & $0.70\pm0.01$ & x & \\    
  $F_{\alpha_2}$& $(1.93\pm0.01)\times 10^{-11}$ & $(6.9\pm1.3)\times 10^{-13}$ & & [$\times 10^{-3}$W m$^{-2}$]=[ergs s$^{-1}$ cm$^{-2}$] \\    
  $r_{\rm in}$& $1.6\pm0.1$ & $1.6\pm0.1$ & x & $r_{\rm g}$\tablenotemark{c} \\    
  $i$ & $39\pm1$& $39\pm1$ & x & degrees \\
  $q_1$ & $3.6^{+0.3}_{-0.6}$ & $3.6^{+0.3}_{-0.6}$ & x &  \\
  $q_2$ & $4.3\pm0.3$ & $4.3\pm0.3$ & x & \\
  $r_{\rm br}$ & $45^{+26}_{-18}$& $> 1.8$ &  & $r_{\rm g}$\tablenotemark{c} \\ 
  $\xi_1$ & $150^{+28}_{-15}$& $38^{+2}_{-1}$ & & [$\times 10^{-3}$W m$^{-2}$]=[ergs s$^{-1}$ cm$^{-2}$] \\
  $F_{\rm{ref_1}}$ & $(3.3\pm0.9)\times 10^{-12}$& $(1.1\pm0.5)\times 10^{-12}$ & & [$\times 10^{-3}$W m$^{-2}$]=[ergs s$^{-1}$ cm$^{-2}$] \\
  $\xi_2$ & $150^{+28}_{-15}$& $38^{+2}_{-1}$ & & [$\times 10^{-3}$W m$^{-2}$]=[ergs s$^{-1}$ cm$^{-2}$] \\
  $F_{\rm{ref_2}}$ & $(3.9\pm0.1)\times 10^{-12}$& $(6.4\pm0.4)\times 10^{-12}$ & & [$\times 10^{-3}$W m$^{-2}$]=[ergs s$^{-1}$ cm$^{-2}$] \\
  $A_{\rm Fe_{1=2}}$ & $1.6\pm0.1$ & $1.6\pm0.1$ & x & solar \\
  \hline
  & Model B & ($\chi2_r = 1.10$)\\      
  \hline
  $\alpha_1$   & $1.64^{+0.04}_{-0.02}$ & $1.64^{+0.04}_{-0.02}$ & x & \\    
  $F_{\alpha_1}$  &  $(4.55\pm0.01)\times 10^{-11}$& $(9.7\pm5.1)\times 10^{-14}$ &
   & [$\times 10^{-3}$W m$^{-2}$]=[ergs s$^{-1}$ cm$^{-2}$] \\    
  $\alpha_2$ &  $0.87\pm0.01$ & $0.74\pm0.01$\\    
  $F_{\alpha_2}$& $(5.87\pm0.09)\times 10^{-12}$ & $(6.27\pm0.63)\times 10^{-13}$ & 
  & [$\times 10^{-3}$W m$^{-2}$]=[ergs s$^{-1}$ cm$^{-2}$] \\    
  $r_{\rm in}$& $< 1.34$&$< 1.34$ & x & $r_{\rm g}$\tablenotemark{c}  \\    
  $i$ & $50\pm1$& $50\pm1$ & x & degrees \\
  $q_{\rm in}$ & $> 4.5$& $5.9\pm0.2$\\
  $q_{\rm out}$ & $>4.5$& $4.4\pm0.2$\\
  $r_{\rm br}$ & $2.8\pm0.1$& $2.8\pm0.2$ & & $r_{\rm g}$\tablenotemark{c} \\ 
  $\xi$ & $20\pm1$& $24\pm1$ & & [$\times 10^{-3}$W m$^{-2}$]=[ergs s$^{-1}$ cm$^{-2}$] \\
  $F_{\rm{ref}}$ & $(1.23\pm0.01)\times 10^{-11}$& $(5.6\pm0.1)\times 10^{-12}$ & & [$\times 10^{-3}$W m$^{-2}$]=[ergs s$^{-1}$ cm$^{-2}$] \\
  $A_{\rm Fe_{1=2}}$ & $0.9\pm0.1$ & $0.9\pm0.1$ & x & solar \\
  \hline   
\enddata 

\tablenotetext{a}{All disk radii are given in units of the gravitational radius and the
outer disk radius is fixed to 400  $r_{|rm g}$. 
The temperature of the MEKAL component is $1.4$~keV. Other parameters listed are 
fluxes $F$,
the inclination angle $i$, the Fe abundance ($A_{Fe}$) given relative to solar, 
 the ionization parameter ($\xi$), and the emissivity index ($q$)}
\tablenotetext{b}{This column marks which model parameters have been tied between the
high and low-state spectra.}
\tablenotetext{c}{Gravitational radius  $r_{\rm g}=GM/c^2$.}

\end{deluxetable}

\begin{deluxetable}{lrcrc}
\tabletypesize{\tiny}
\tablecaption{Measurements of the optical emission lines
\label{line_summary}}
\tablewidth{0pt}
\tablehead{
\colhead{Line} & 
\colhead{EW\tablenotemark{1}} &
\colhead{$\lambda_{\rm c,obs}$\tablenotemark{2}}  &
\colhead{FWHM\tablenotemark{3}} &
\colhead{x/[OIII]\tablenotemark{4}}
} 
\startdata
He II4686n   & 5.0 & 4807  & 950 & 0.19 \\
He II4686b  & 16.9 & 4799  & 5100 & 0.59 \\
H$\beta$n  & 4.0 & 4984.2 & 560 & 0.13 \\
H$\beta$b  & 90.0 & 4984.2 & 1700 & 3.23 \\
$\rm [OIII]$4959  & 8.4 & 5086.7  & 500 & 0.31 \\
$\rm [OIII]$5007  & 28.1 & 5136.1 & 500 & 1.00 \\
$\rm [Fe VII]$5159  & 3.4 & 5293.3  & 1250 & 0.11 \\
$\rm [Fe XIV]$5303  & 2.8 & 5441.3 & 1350 & 0.09 \\
$\rm [Fe VII]$5721  & 4.0 & 5866.7  & 1250 & 0.14 \\
He I 5876  & 15.8 & 6025.7 & 1420 & 0.54 \\
$\rm [Fe VII]$6087  & 3.0 & 6239.6  & 930 & 0.09 \\
$\rm [Fe X]$6375  & 3.9 & 6535.8   & 900 & 0.13 
\enddata

\tablenotetext{1}{The equivalent widths are given in units of \AA.}
\tablenotetext{2}{The observed central wavelength of a line is given in \AA.}
\tablenotetext{3}{Full-width at half-maximum of the line given in units of km s$^{-1}$.}
\tablenotetext{4}{{Line} fluxes relative to [OIII]$\lambda$5007 line flux 
($F_{\rm [OIII]}=1.93\times 10^{-16}$ W m$^{-2}$}

\end{deluxetable}

\clearpage

\begin{figure*}
\epsscale{1.2}
\plottwo{f7aa.ps}{f7ba.ps}

\plottwo{f7ab.ps}{f7bb.ps}

\plottwo{f7ac.ps}{f7bc.ps}
\caption{\label{mkn335_2006_2007_ref}
Fits to the 2006 (high-state) and 2007 (low-state) \xmm\ pn spectra with reflection models. 
The left panels display the fits using model A (i.e. two power law components
both with associated reflection) and the right panels for model B (i.e. two power law
components, of which only one (the flatter) with associated
reflection). The upper panels display the 
fits to the high-state, the middle panels the low state, and the bottom panels the residuals of the model fits. 
The line spectra (mekal + emission lines) are in black,
power laws are shown in red,
 both reflection components are in blue,
and the total model is green.
The spectral parameters of the joint fits 
in the 0.3-12.0 keV band are given in Table\,\ref{reflection_model}.
} 
\end{figure*}

\begin{figure}
\epsscale{0.8}
\plotone{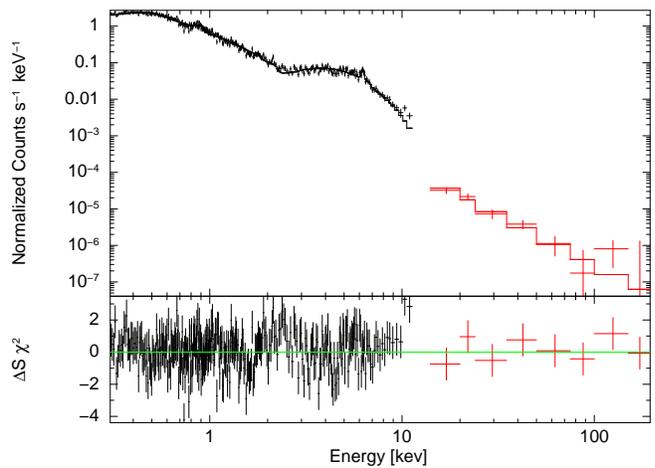}
\caption{\label{mkn335_xmm_bat}
Combined \xmm\ pn and \swift\ BAT spectrum fitted with a power law with partial
covering absorber. 
} 
\end{figure}

\begin{figure*}
\epsscale{0.8}
\plotone{f9a.ps}
\epsscale{1.2}

\plottwo{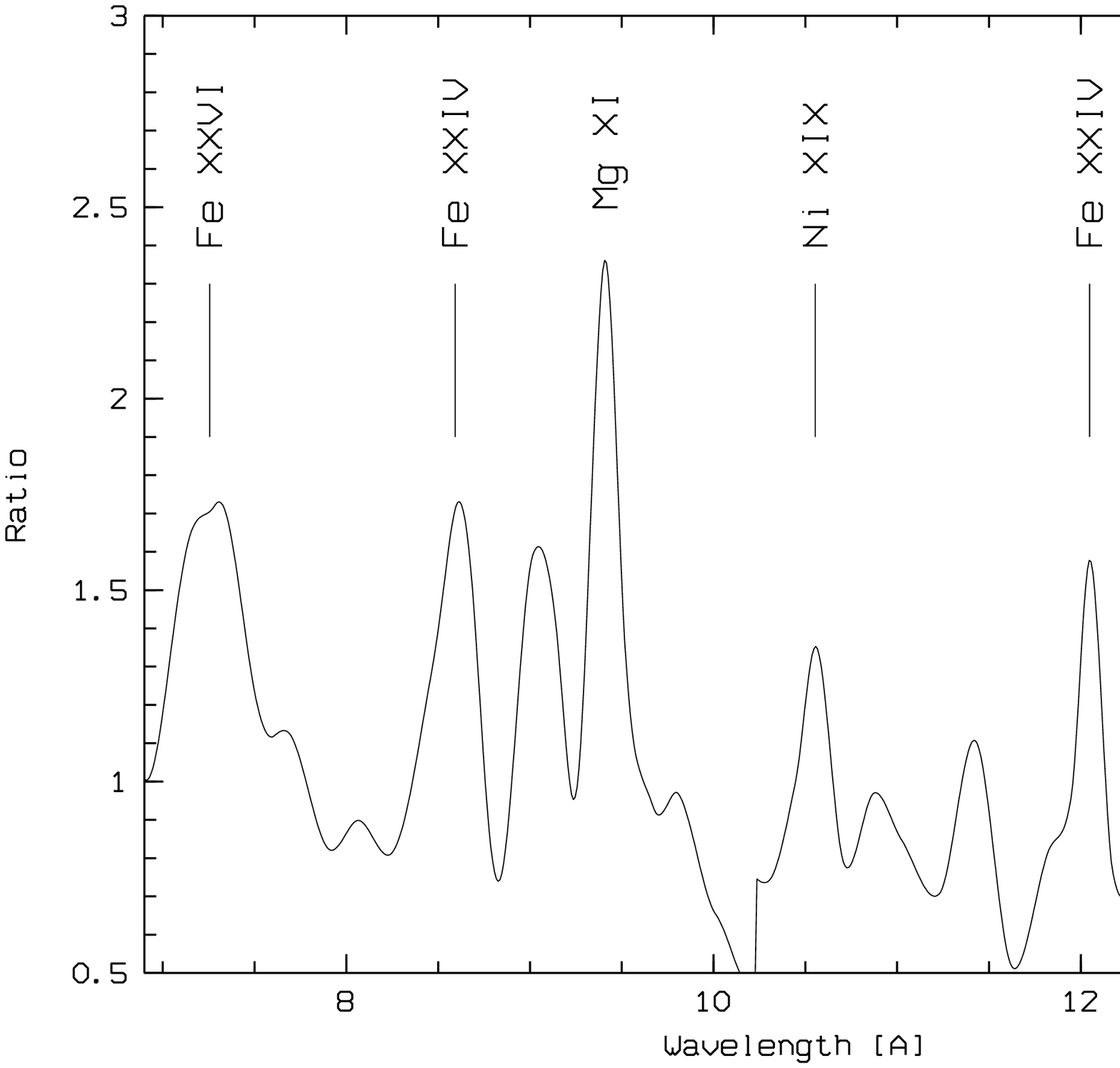}{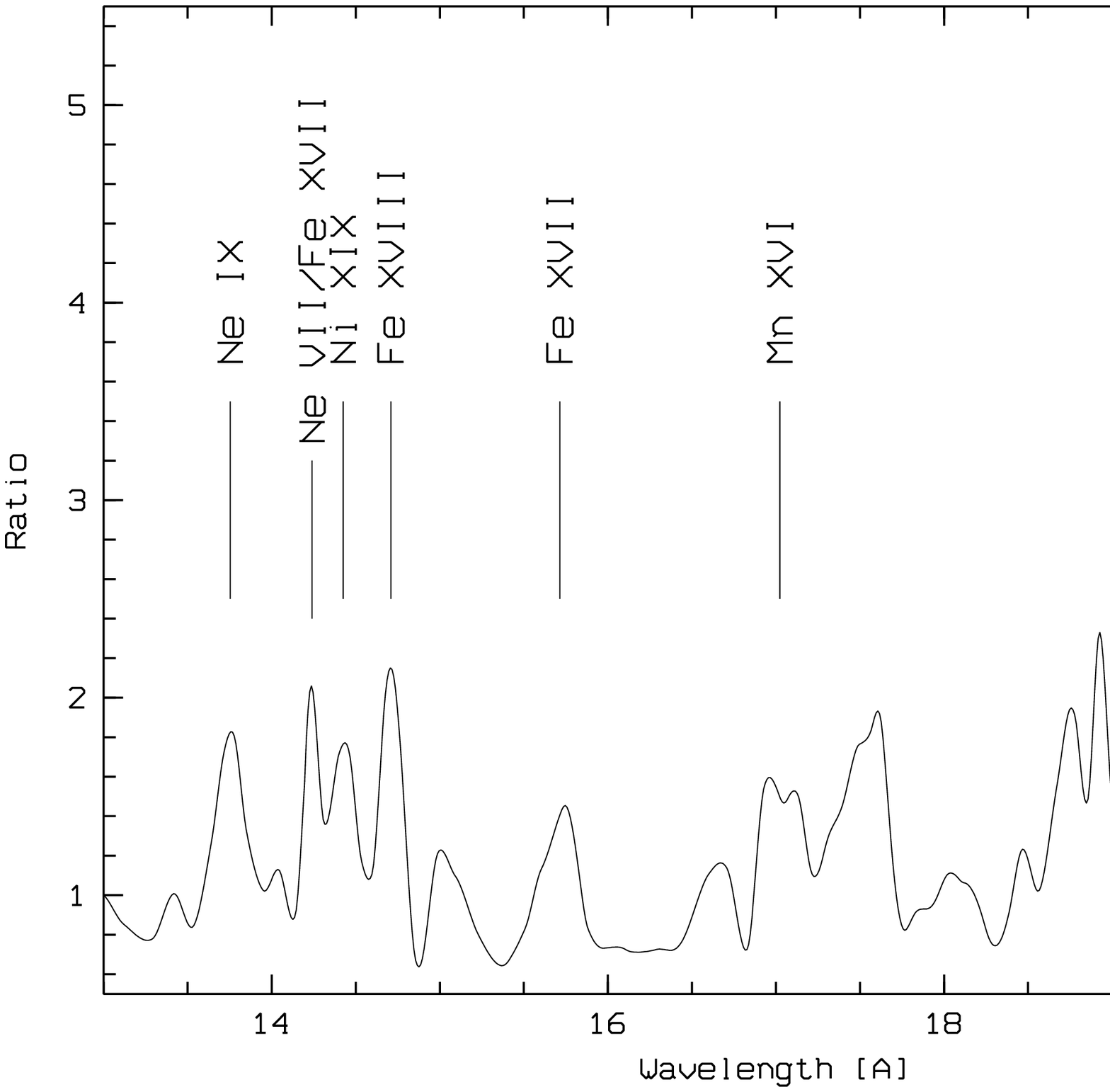}

\plottwo{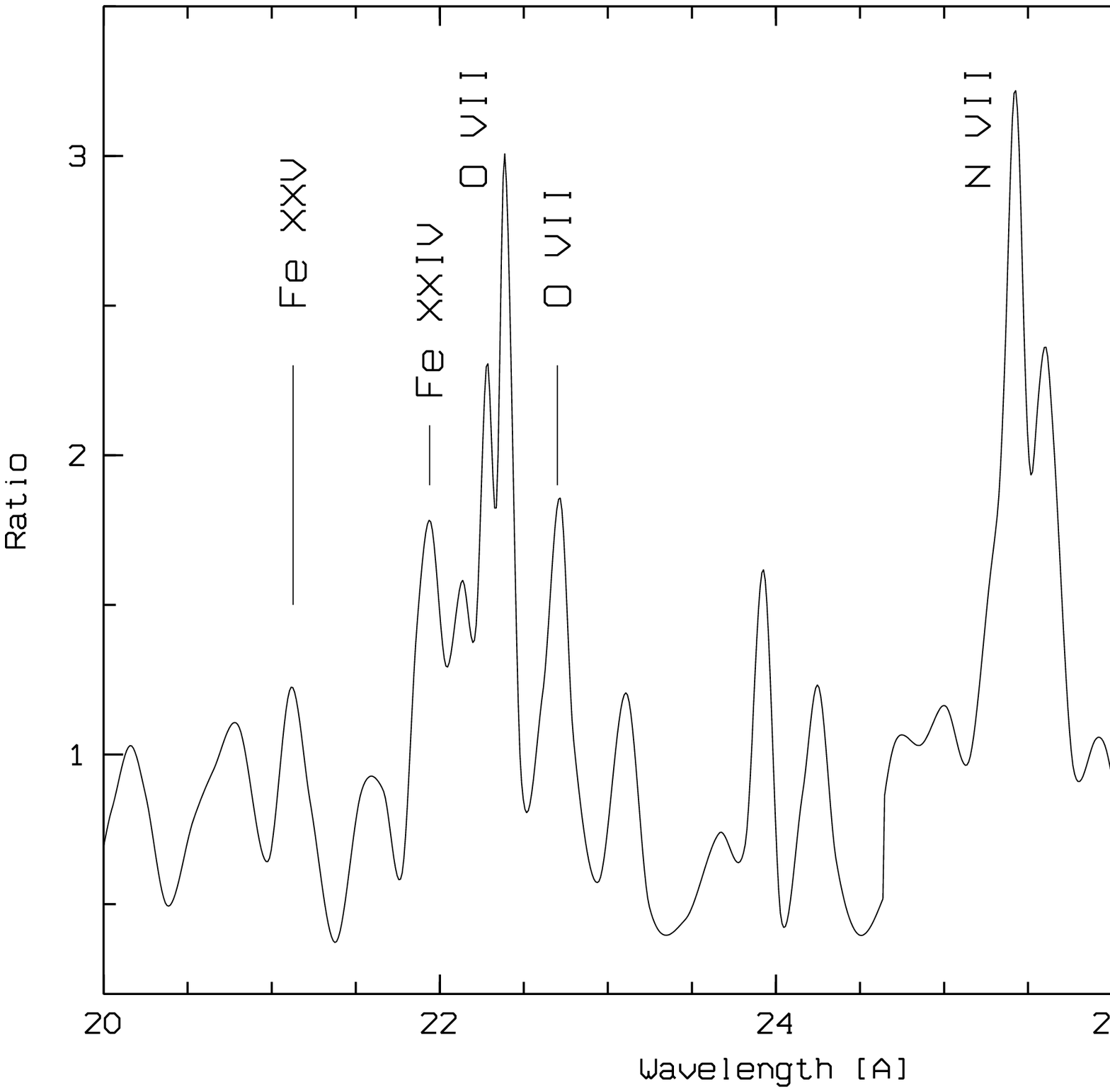}{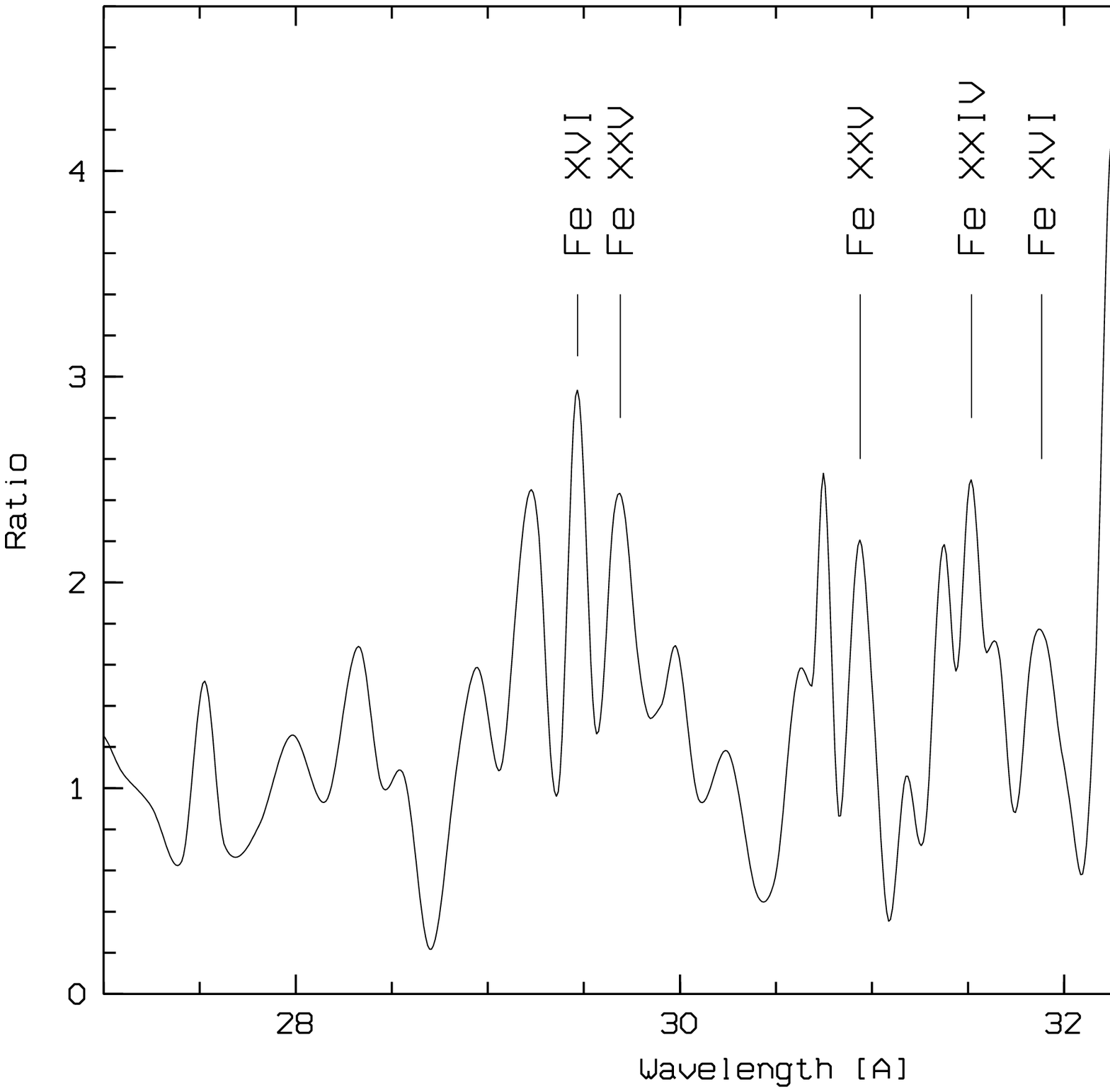}
\caption{\label{mkn335_rgs} 
RGS spectra of Mrk 335. The upper panel displays the RGS spectra fitted by a single power
law model. Clearly the residuals reveal several soft X-ray emission lines. 
The lower panels display 
the theoretical positions of some expected
  emission lines which are marked in the spectrum at different wavelengths ranges.  
}
\end{figure*}

\begin{figure*}
\epsscale{1.0}
\plotone{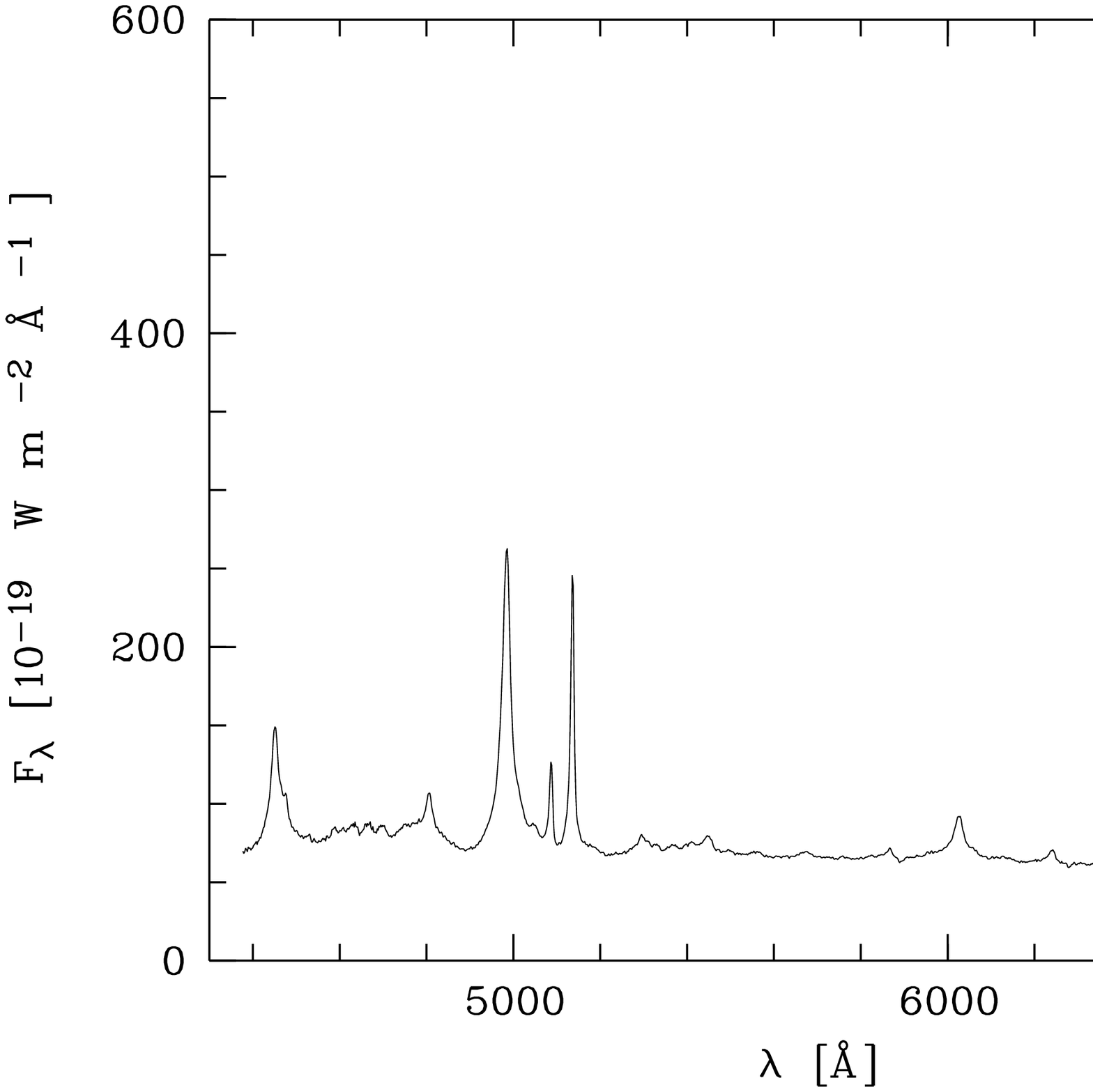}
\epsscale{1.2}

\plottwo{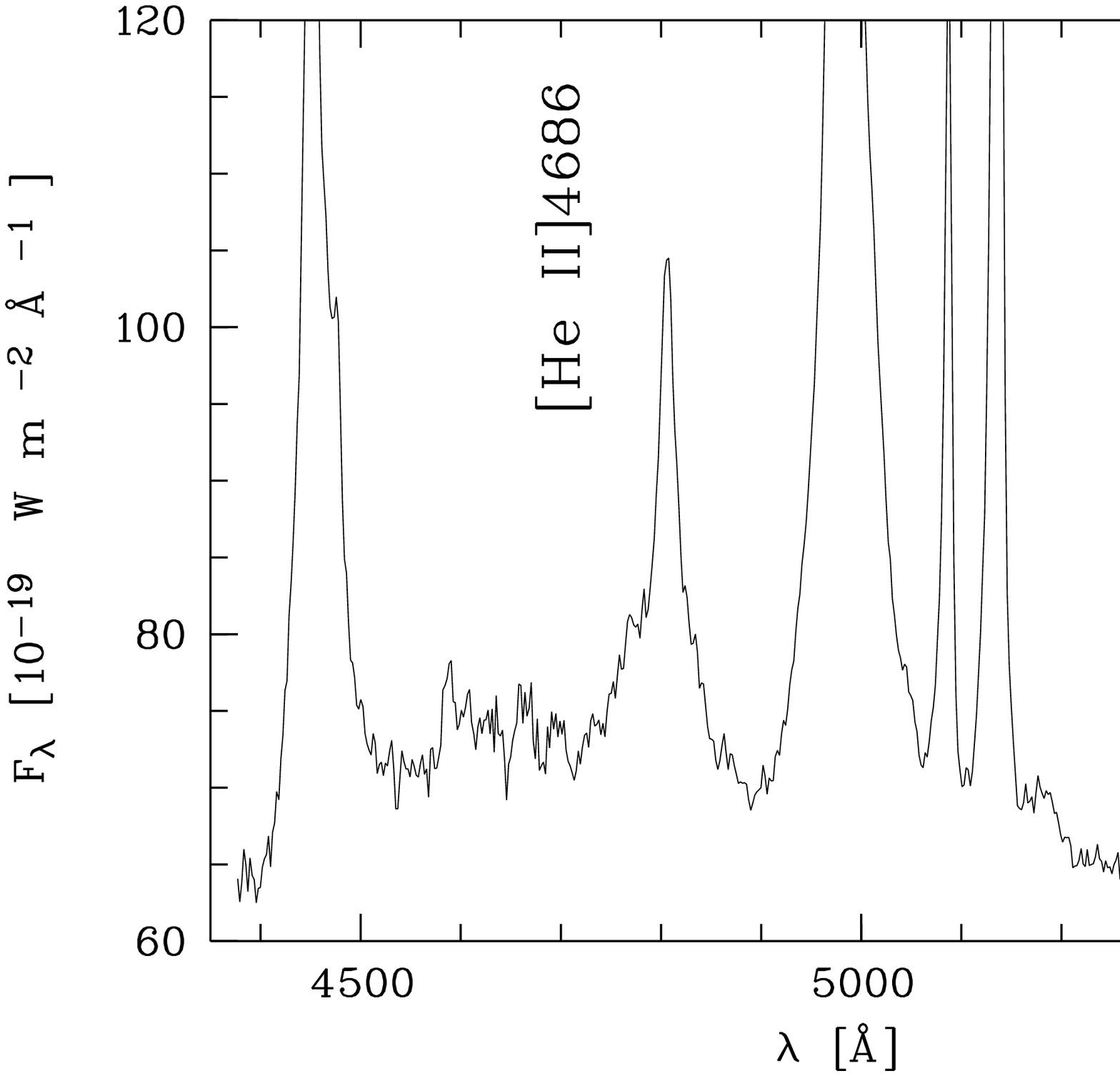}{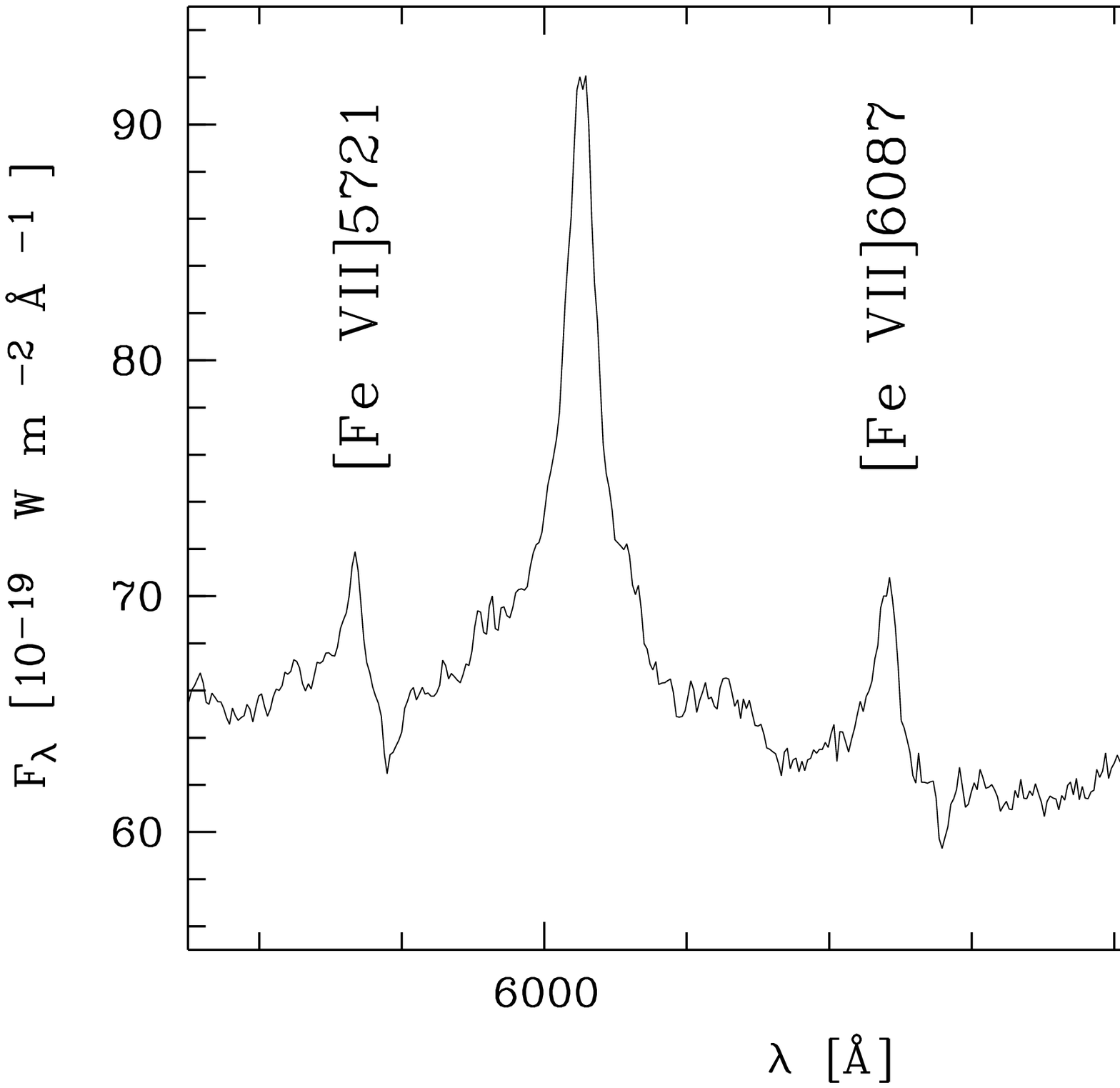}
\caption{\label{mkn335_fe} 
Optical spectrum of Mrk 335. The upper panel displays the whole optical spectrum taken on
2007 September 07. The lower two panels show the
 blue and red regions of the highly ionized coronal Fe 
line in the FeII-subtracted optical spectrum of Mrk 335. 
}
\end{figure*}

\clearpage



\end{document}